\documentclass[useAMS,usenatbib,usegraphicx,referee]{mn2e}

\title[Simulation of a population of neutron stars...]{Simulation of a population of isolated neutron stars evolving through the emission of gravitational waves}
\author[C. Palomba]{C. Palomba$^{1}$\thanks{E-mail:cristiano.palomba@roma1.infn.it}\\
$^{1}$Istituto Nazionale di Fisica Nucleare, sezione di Roma\\
P.le A. Moro, 2 - 00185 Roma, Italy}

\begin{document}

\date{}
\pagerange{\pageref{firstpage}--\pageref{lastpage}} \pubyear{}

\maketitle
\label{firstpage}

\begin{abstract}
We study, via a Monte Carlo simulation, a population of isolated asymmetric neutron stars where the magnitude of the magnetic field is low enough so that the dynamical evolution is dominated by the emission of gravitational waves. A starting population, with age uniformly distributed back to 100 Myr (or 500 Myr) and endowed with a birth kick velocity, is evolved in the Galactic gravitational potential to the present time. In describing the initial spatial distribution, the Gould Belt, with an enhanced neutron star formation rate, is taken into account. Different models for the initial period distribution are considered. The star ellipticity, measuring the amount of deformation, is drawn from an exponential distribution. 
We estimate the detectability of the emitted gravitational signals by the first and planned second generation of interferometric detectors.
Results are parametrized by the fraction of the whole galactic neutron star population made of this kind of sources.
Some possible mechanisms, which would make possible the existence of such a population, are discussed. A comparison of the gravitational spin-down with the braking due to a possible interaction of the neutron star with the interstellar medium is also presented.
\end{abstract}


\section{Introduction} \label{intro}
\begin{figure}
\includegraphics[width=84mm]{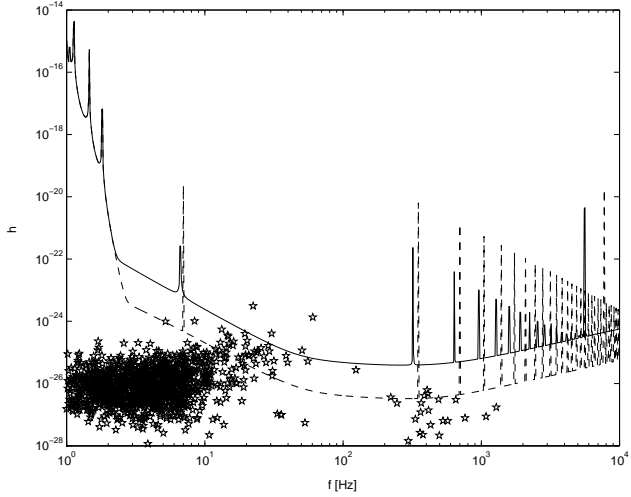}
%%%\vspace{2cm}
\caption{Effective amplitude of the gravitational wave signals emitted by known isolated pulsars versus the target sensitivity of the Virgo detector (continuous line) and a possible sensitivity curve for an advanced Virgo detector (dashed line). The total observation time is four months and the use of an {\em optimal} data analysis method is assumed. The observed spin-down is assumed to be due completely to the emission of gravitational radiation.}
\label{pulsar1}
\end{figure}

Continuous gravitational signals emitted by rotating asymmetric neutron
stars are promising sources for interferometric gravitational wave detectors like GEO \citep{geo}, LIGO \citep{ligo}, TAMA300 \citep{tama} and VIRGO \citep{virgo}. The
detectability of a source, emitting due to a deviation of its shape from axi-symmetry, basically depends on the detector sensitivity, on the source rotation
frequency, distance and ellipticity. Not less important for the
feasibility of the search are, however, also the source position in the
sky and spin-down rate. Up to now about 1500 neutron stars, most of which are pulsars, have been
detected in the Galaxy through their electromagnetic emission. Standard theory of
stellar evolution suggests that about $10^9$ neutron stars should exist in
the Galaxy and, of these, about $10^5$ should be  
active pulsars. We detect only a small fraction of this subset because
either the radio beam does not intersect our visual line or the intensity
of the emission is too low. In fact, the sample of detected pulsars is
increasing as the sensitivity of the surveys increases. For most of these
stars the rotation frequency, the spin-down rate, the location in the sky
and the distance are known, while the ellipticity is completely uncertain.
This makes difficult to get reliable estimates of their gravitational
emission. An upper limit to the ellipticity, and then to the amplitude of 
the emitted gravitational signals, is tipically given assuming that all the observed spin-down is due
to the emission of gravitational waves. In this case $\epsilon\le 1.8\cdot 10^5\sqrt{\dot{P}P^3}$, being $P,\dot{P}$ respectively the rotation period and its first time derivative and, for most normal pulsars, values of the order of $10^{-3}\div 10^{-2}$ are obtained.
Using this assumption, in Fig.(\ref{pulsar1}) we show the effective amplitude of the gravitational wave signals emitted by the presently known isolated pulsars\footnote{http://www.atnf.csiro.au/research/pulsar/psrcat/. We do not take into account pulsars in binary systems for which the detection would be complicated by the system orbital Doppler effect and by the stochastic fluctuations of the rotation period due to the accretion of matter.}, after using {\em optimal}
filtering techniques and four months of observation time, versus the target sensitivity of the Virgo detector and a possible sensitivity curve of an advanced Virgo detector \citep{punt}. Similar plots can be obtained for the LIGO interferometers. It comes out that very few observed pulsars would be on the verge of detectability, for the first generation of interferometers. Tens of pulsars would be detected by an advanced detector.  
\begin{figure}
\includegraphics[width=84mm]{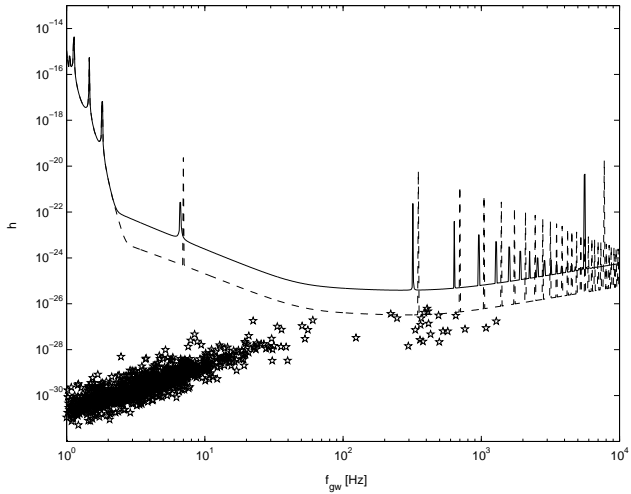}
%%%\vspace{2cm}
\caption{Same as Fig.(1) but taking the source ellipticity as the minimum between $10^{-6}$ and that obtained assuming the observed spin-down is completely due to the emission of gravitational radiation.}
\label{pulsar2}
\end{figure}
On the other hand, as it is well known, it is quite likely that true ellipticities are much lower than the upper limits. According to \citet{usho} the maximum ellipticity is probably $\la 10^{-6}$. 
This makes rather unlikely the detection of gravitational waves from the observed pulsar population: Fig.(\ref{pulsar2}) is the same as Fig.(\ref{pulsar1}) but taking the ellipticity as the minimum between $10^{-6}$ and the value obtained assuming only gravitational spin-down. The observed spin-down of millisecond pulsars implies a maximum ellipticity less than $~10^{-8}$, which is a plausible value. In this case we see that no detection is expected for the first generation of interferometers while few sources (4 at present), all millisecond recycled pulsars, could be detected by an advanced detector. By the way, also this is just an upper limit.  
Nevertheless, the search for gravitational signals from known neutron stars is an important task because, even if no detection should come, we can at least put upper limits on the neutron stars ellipticity, thus increasing our knowledge on the neutron star structure.

An alternative possibility is the existence of a population of asymmetric
rotating neutron stars with magnetic field low enough so that
the spin-down is dominated by the emission of gravitational waves. For such 
objects, as the electromagnetic emission does not contribute much or at
all to the
spin-down, a larger amount of gravitational radiation would be emitted in a
given period of observation, in comparison to a standard pulsar with the same degree of deformation. Obviously, it would be very difficult to detect this kind of neutron stars in the electromagnetic
band, unless very near and young (through their thermal emission) or very near and interacting with the interstellar medium. Hence the necessity of performing a {\em blind search} in the data of gravitational detectors, i.e. without strong assumptions on the position of the objects, on their rotation frequency and spin-down rate. The existence of such class of stars, which for simplicity we
call GWDNS (Gravitational Wave Driven Neutron Stars) or {\em gravitars}, cannot be excluded both on
observational and theoretical basis. For instance, the lack of a
one-to-one correspondence between observed supernovae and observed compact
remnants leaves room for a non negligible formation rate for GWDNS. In this context, the presence of a low magnetic field neutron star has been recently suggested for the remnants of $SN~1987~A$ \citep{ogel}. 
GWDNS could be a subset of pulsars born with a
very low magnetic field or the product of a different evolutionary channel.   
Different mechanisms, like magnetic axis alignement, the freezing of the magnetic field in a newborn neutron star or magnetic field decay, could permit the existence of GWDNS.

In this work we perform a Monte Carlo simulation of a possible
GWDNS population, estimating the
number of expected detections with the first and planned second generation of gravitational
interferometers. 
Several simulations of the neutron star or pulsar population, evolving through standard dipolar magnetic emission, have been published \citep{pacz,nara,lyne,hans,arzo,gonth}. Only in few cases the detection of gravitational waves has been discussed
\citep{regi}. The existence of a population of neutron stars whose dynamical evolution is dominated by the emission of gravitational waves has been addressed by \citet{bland} but, as far as we know, no attempt has been done, up to now, to make a quantitative study of the problem. We do not have an observed population
against which the simulated one can be checked so that different choices for
the parameters are done and the dependency of the 
results on them is discussed. The initial position and kick velocity are described by the probability distributions typically used to simulate the standard pulsar population. The contribution to the total neutron star formation rate due to the Gould Belt is taken into account and three different distributions for the initial rotation period are considered.
The mean value of the ellipticity, distributed exponentially, is taken between $10^{-8}$ and $10^{-6}$ and the abundance of GWDNS, expressed as fraction of the whole galactic neutron star population, is considered as a free parameter.  
Given the initial spatial, initial period and birth kick velocity distributions, GWDNS are evolved solving the equations of motion in the galactic
gravitational potential, assumed axisymmetric for simplicity. The stars age in drawn from a uniform distribution back to 100 Myr (or 500 Myr, when advanced detectors are considered).
Once the population has been evolved to the present, the detectability of the emitted continuous gravitational signals is estimated, for the first and planned second generation of interferometric detectors, assuming that a {\em blind} search is done. This is mandatory because, contrary to what happens in {\em targeted} search where the source parameters are well constrained, here we need to explore a large portion of the source parametrer space.
This implies that {\em optimal} data analysis methods cannot be applied, due to the unrealistic amount of computing resources they would require. Non optimal procedures, with reduced sensitivity but allowing to strongly cut the computational weigth of the analysis, are therefore needed.

The plan of the paper is the following. In Section 2 we review the main relations concerning the time evolution of neutron stars rotation period due to the presence of a dipolar magnetic field or of an asymmetry in the star shape. In Section 3 the sensitivity of a {\em blind} search for the gravitational waves emitted by asymmetric rotating neutron stars is given. In Section 4 the Monte Carlo simulation model is presented and in Section 5 the results of the simulations are discussed. In Section 6 we describe some possible mechanisms which could leave space to a non negligible fraction of GWDNS. In Section 7 conclusions are drawn. Finally, in the Appendix we compare the gravitational spin-down of a neutron star with the spin-down due to its possible interaction with the interstellar medium.

\section{Spin evolution basic relations} \label{spin}
The spin-down rate of a neutron star can be expressed, in the general case, as
\begin{equation}
\dot{\Omega}=-\sum_{i} K_i\Omega^{n_i}
\label{omegadot}
\end{equation}
where $\Omega$ is the rotation pulsation, $n_i$ are the {\em braking indexes} which depend on the spin-down mechanisms at work and $K_i$ are constants depending on the neutron star structure and characteristics.
In the case of simple electromagnetic dipolar emission  
Eq.(\ref{omegadot}) becomes
\begin{equation}
\dot{\Omega}_{{dip}}=-{2\over{3c^3}}{B^2\over I}R^6\Omega^3\sin{\alpha}
\label{omegadotem}
\end{equation}
where $B$ is the magnitude of the dipolar magnetic field, $\alpha$ is the angle between the magnetic axis and the rotation axis, $R$ is the neutron star radius and $I$ is the star momentum of inertia.
If the evolution is dominated by the emission of gravitational waves in a non-axysimmetric neutron star, we have
\begin{equation}
\dot{\Omega}_{{gw}}=-{32\over 5}{G\over c^5}I\epsilon ^2 \Omega^5
\label{omegadotgw}
\end{equation}
where $\epsilon={{I_1-I_2}\over I_3}$ is the star ellipticity, being $I_1,I_2,I_3$ the star momenta of inertia along three principal axes. 
From this equation the rotation frequency as a function of time can be expressed as
\begin{equation}
f(t)=\left(P_0^4+{{2048\pi^4}\over 5}{G\over c^5}I\epsilon^2t\right)^{-{1\over 4}}
\label{ftime}
\end{equation}
where $P_0$ is the initial rotation period.
The final frequency is independent on the initial period if the inequality
\begin{equation}
P_0\ll 0.0443\left({I\over{1.12\cdot 10^{38}kg~m^2}}\right)^{1\over 4}\left({\epsilon\over 10^{-6}}\right)^{1\over 2}\left({t\over {1~Gyr}}\right)^{1\over 4}~~s
\label{P0}
\end{equation}
is satisfied and, in such a case, the rotation frequency at present is
\begin{equation}
f(t)\simeq 23 \left({I\over{1.12\cdot 10^{38}kg~m^2}}\right)^{-{1\over 4}}\left({\epsilon\over 10^{-6}}\right)^{-{1\over 2}}\left({t\over {1~Gyr}}\right)^{-{1\over 4}}~~Hz
\label{f0}
\end{equation}
We note that even very old and rather highly deformed GWDNS can continue to spin to a rather high frequency at present. GWDNS only slightly deformed can maintain nearly unchanged their rotation frequency along their whole life.

We are interested in comparing the contributions to the spin-down rate due to the emission of gravitational waves and to the presence of a dipolar magnetic field. 
In this way we can establish which condition the magnetic field must satisfy in order to have a GWDNS.
Let us introduce \citep{palo} the ratio among the gravitational and the dipolar field induced spin-down rates, expressed by Eqs.(\ref{omegadotgw},\ref{omegadotem}):
\begin{equation}
Y={\dot{\Omega}_{{gw}}\over{\dot{\Omega}_{{dip}}}}=2.3\cdot 10^{-7}\left({I\over{1.12\cdot 10^{38}kg~m^2}}\right)^2\left({\epsilon \over 10^{-6}}\right)^2\left({f\over {100 Hz}}\right)^2\left({B\over {10^{12}G}}\right)^{-2}\left({R\over {10km}}\right)^{-6}\sin{\alpha}^{-2}
\label{Y}
\end{equation}
The condition $Y> 1$ corresponds to
\begin{equation}
B< 4.75\cdot 10^8\left({I\over{1.12\cdot 10^{38}kg~m^2}}\right)\left({\epsilon\over 10^{-6}}\right)\left({f\over {100 Hz}}\right)\left({R\over {10km}}\right)^{-3}\sin{\alpha}^{-1}~~G
\label{B}
\end{equation}
Obviously, this relation is less stringent for high values of the neutron star ellipticity and frequency and small values of the angle $\alpha$.
In order to have a GWDNS is not sufficient that the condition $Y>1$ is satisfied at the birth. Due to the spin-down, the ratio $Y$, even if $>1$ at the beginning, decreases in time and could go below $1$, unless the magnetic field decays or the magnetic axis aligns to the rotation axis. These possibilities will be discussed in Section 6.
Assuming that the "non-gravitational" spin-down is described by Eq.(\ref{omegadotem}), we are implicitly considering a neutron star in the so-called {\em ejector} phase \citep{popov}. This is the evolutionary stage to which pulsars belong. Note, however, that not all {\em ejectors} are pulsars. Two other phases are foreseen for neutron stars having small values of the magnetic field and/or long rotation periods: {\em propeller} and {\em accretor}. In these phases an interaction of the star with the sorrounding interstellar medium takes place and in the Appendix we will see that, unless the magnetic field is nearly vanishing, the spin-down rate in these stages is dominated, for $\epsilon \ga 10^{-9}$, by the emission of gravitational waves.

\section{Detection of continuous gravitational signals} \label{detect}
The gravitational wave signal emitted by an asymmetric neutron star, rotating around one of its principal axis of inertia, has a frequency twice the rotation frequency and the amplitude at the detector, assuming optimal orientation, can be expressed as
\begin{equation}
h_0=1.05\cdot 10^{-27}\left({I\over{1.12\cdot 10^{38}kg~m^2}}\right)\left({\epsilon \over 10^{-6}}\right)\left({f_{gw}\over {100 Hz}}\right)^2\left({{10~kpc}\over r}\right)
\label{h0}
\end{equation}
where $f_{gw}={2\over P}$ is the gravitational wave frequency and $r$ the distance from the detector. 
From Eq.(\ref{ftime}), assuming $\epsilon \le 10^{-6}$, most GWDNS emit gravitational waves with frequency $\ga 10~Hz$, independently of their age, i.e. within the sensitivity band of interferometric detectors.
However, we will see in Section \ref{res} that almost all GWDNS producing signals detectable by the first generation of interferometric detectors are younger than $10~Myr$.
The expected amplitude of the gravitational signals is so low that they will be typically buried in the detector noise. Then, very long integration times (of the order of months) and suitable data analysis techniques are needed to extract them. 
It is important to take into account that the signal "seen" by the detector is not really monochromatic due to the source spin-down and the Doppler effect associated to the detector motion. Moreover,    
the search for GWDNS must be necessarily {\em blind}, that is it must be performed on a large portion of the source parameter space, because we cannot put strong constraints on their position in the sky, on their rotation frequency and on their spin-down rate. This means, in practice, that we want to search sources everywhere on the celestial sphere ({\em full-sky} search), with signal frequencies from the low frequency cut-off of interferometers up to about $2~kHz$, and a spin-down rate as large as possible, compatibly with the available computing resources.
Such kind of analysis cannot be performed with the classical {\em optimal} methods used in gravitational wave experiments, based on the so-called {\em matched filter}, which can be used for {\em targeted} searches \citep{rogpulsar,ligopulsar}, but that are unfeasible for wide area searches, due to the enormous computing power they would require. Alternative, non optimal, methods have been developed by various groups in the world \citep{fras1,ast,brad,schu,fras2}. They reach a lower level of sensitivity but strongly cut the needed computational power. All these methods are based on some kind of alternation of coherent and incoherent steps, these last typically based on the Radon \citep{brad} or Hough transform \citep{fras2,kris}. 
In the hierarchical method developed by our group \citep{fras2,broc,fras3} the amplitude of the smallest detectable signal, with signal-to-noise ratio $\sim 2$, is
\begin{equation}
h_{{min}}(f)=\left({4S^2_h(f)\over{T_{{obs}}T_{{FFT}}}}\right)^{1\over 4}\beta
\label{sensi}
\end{equation}
where $S_h(f)$ is the detector noise power density, measured in $Hz^{-1}$, $T_{obs}$ is the total observation time, and $T_{{FFT}}$ is the time duration of the {\it short} FFTs computed from the data at the beginning. $T_{{FFT}}$ varies between $1048~s$ and $16667~s$, depending on the frequency band which is analyzed. The {\em loss factor} $\beta$ is a number around 2, which exact value depends on the analyzed frequency band, on the number of candidates which are selected after the first incoherent step and on the assumed minimum spin-down decay time.
The larger is the available computing power and the wider can be the search. 
A computing power of about 1 Teraflops will allow to perform a {\em full-sky} analysis of $4$ months of data, searching sources emitting signals with frequencies up to $2~kHz$ and with a spin-down time $\tau={f\over {\dot{f}}}>10^4~yr$.

\begin{figure}
\includegraphics[width=84mm]{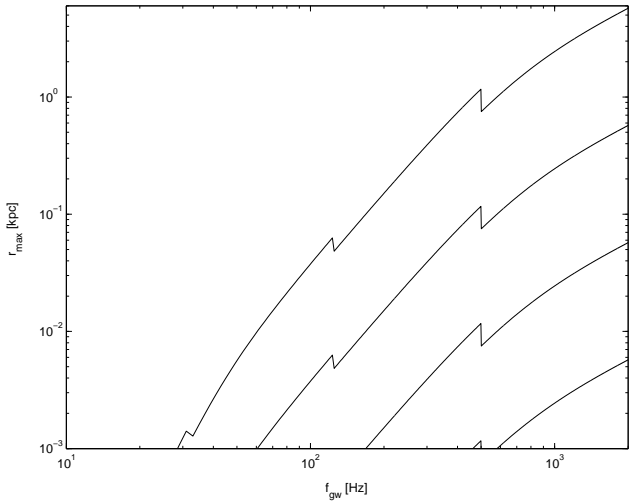}
%%\vspace{2cm}
\caption{Maximum distance at which the gravitational signal emitted by an asymmetric rotating neutron star could be detected, as a function of the signal frequency, assuming the hierarchical procedure developed by our group is applied and using the Virgo target sensitivity. The search parameters are: total observation time of four months, minimum spin-down decay time of $10^4~yr$, $10^9$ candidates selected after the first incoherent step. The different curves corresponds to different values of the star ellipticity, starting from $10^{-6}$ (upper curve) to $10^{-9}$ (lower curve). \label{rmax}}
\end{figure}

From Eqs.(\ref{h0},\ref{sensi}) we can derive the maximum distance $r_{max}$ at which a source, emitting gravitational waves at a given frequency and with a given ellipticity, can be detected.  
In Fig.(\ref{rmax}) $r_{max}$ is plotted as a function of the frequency for different values of the ellipticity. A total observation time of $4$ months is considered and the target Virgo sensitivity is used. The discontinuities are due to the fact that different values of the $FFT$ duration, $T_{FFT}$, are used for different frequency bands.
Looking at the figure we see that if $\epsilon \le 10^{-6}$ and if the maximum rotation frequency we search for is $1~kHz$, the maximum distance at which a source can be detected, in a {\em blind} search, is about $6~kpc$. 
On the other side, assuming that no neutron star is located at a distance from the Earth less than $10~pc$, the minimum ellipticity needed to have a detectable source (emitting at the high end of the allowed frequency range) is $\sim 1.4\cdot 10^{-9}$: no source can be detected, not even in principle, if its ellipticity is less than that value. If the typical ellipticity is in the range $10^{-8}\div 10^{-7}$ we would need to have sources located in the solar neighbourhood and rotating at high rate. Moreover, sources with rotation frequency less than $\sim 25~Hz$ cannot be detected, indipendently of the ellipticity. In Section 5 we will see that detected sources in the simulation have parameters in a narrower range.

\section{Monte Carlo simulation model} \label{monte}
In this section we describe the model used to simulate the population of GWDNS in the Galaxy. As we do not have any observational clue on what the parameters distribution of GWDNS could be, we use for initial position and kick velocity the same distributions typically used for pulsars. Three different models are considered for the distribution of initial periods. The ellipticity follows an exponential distribution with mean value treated as a parameter. Obviously we cannot exclude that GWDNS have parameters distributed in a completely different way. We take the distance of the Sun from the galactic centre as $R_\odot =8.5~kpc$. 
 
\subsection{Space distribution} \label{spa}
We assume that the initial space distribution is described by two exponentials in the radial distance from the galatic centre, $R=\sqrt{x^2+y^2}$, and in the height over the galactic plane $z$:
\begin{equation}
p(R)={1\over R_0}e^{-{R\over R_0}}
\label{rdistr}
\end{equation}
with scale factor $R_0=3.2~kpc$, and
\begin{equation}
p(z)={1\over {2z_0}}e^{-{|z|\over z_0}}
\label{rdistr}
\end{equation}
with $z_0=0.075~kpc$. The results of the simulations are rather insensitive to the choice of the scale factors. 
The initial right ascension, $\phi_0$, is chosen from a uniform distribution between 0 and $2\pi$. The angle is measured respect to the direction between the galactic centre and the Sun.
With these choices we find that about $1100$ neutron stars were born in the galactic disk up to a distance of $3~kpc$ from the Sun in the last $4~Myr$ (assuming a neutron star formation rate of $0.02yr^{-1}$), in good agreement with \citet{pop}.
The presence of the Gould Belt is also taken into account. It is described as a thin disk with radius of $300~pc$ and thickness $60~pc$, tilted of $18^o$ respect to the galactic plane and centered at $100~pc$ from the Sun in the Galactic anticenter direction \citep{pop}. The Gould Belt is characterized by an overabundance of massive stars and its age is about $40~Myr$. Following \citet{pop}, we assume a constant star formation rate inside the belt of $27$ neutron star per Myr. To mimic the observed lack of massive stars in the central region of the Gould Belt, we assume that no neutron star was born in its central region up to a distance of $150~pc$ from the center \citep{popov2}. We find that a fraction $\sim 1.5\cdot 10^{-3}$ of the neutron star population with age less than $40~Myr$ was formed in the solar neighbourhood, at a distance from the Sun less than $600~pc$, and this is in agreement with the result of \citet{popov2}.

\subsection{Kick velocity distribution}
We assume that the distribution of the modulus of the kick velocity $v_{kick}$ imparted to neutron stars at birth is  that given by \citet{arzo}:
\begin{equation}
p(v_{{kick}})=4\pi v^2_{{kick}}\left({w_1\over{\left(2\pi \sigma^2_{{v_1}}\right)^{3\over 2}}}e^{-{v^2_{{kick}}\over{2\sigma^2_{{v_1}}}}}
+{{(1-w_1)}\over{\left(2\pi \sigma^2_{{v_2}}\right)^{3\over 2}}}e^{-{v^2_{{kick}}\over{2\sigma^2_{{v_2}}}}}\right)
\label{vdist}
\end{equation}
where $w_1=0.4$, $\sigma_{{v_1}}=90~km s^{-1}$, $\sigma_{{v_2}}=500~km s^{-1}$.
The angles $\phi_v,\theta_v$, defining the direction of the kick velocity vector, are drawn from uniform distributions between, respectively, $[0,2\pi)$ and $[-{\pi\over 2},{\pi \over 2}]$. 

\subsection{Initial period distribution} \label{initper}
We use three different models for the initial period distribution.
The first one is a {\em standard} log-normal distribution 
\begin{equation}
p(P_0)={1\over{\sqrt{2\pi}\sigma_{\log{P_0}}P_0}}e^{-{(\log{P_0}-\overline{\log{P_0}})^2}\over{2\sigma^2_{\log{P_0}}}}
\label{p0dist}
\end{equation}
with $\overline{\log{P_0}}=-2.3$ and $\sigma_{\log{P_0}}=0.3$ \citep{arzo}. 
GWDNS with $P_0<0.5~ms$ are excluded from the simulation. 
The second model is like the first one but we set to $10~ms$ all initial periods shorter than this value. In this way we try to mimic the possible presence of excited {\em r-modes} \citep{owen} which could rapidly increase the rotation period of a fast rotating newborn neutron star. According to \citet{ander}, see section 3 of their paper, the final period of a newborn neutron star, once the {\em r-modes} have become stable, 
is between $\sim 5~ms$ and $\sim 15~ms$, depending on the cooling scenario. We take $10~ms$ as a intermediate value. 
The third model consists of a uniform distribution of periods between $2~ms$ and $15~ms$, mimicing in this way the 
combined effect of {\em r-modes}, which tend to increase the rotation period, and of the matter fall-back after supernova 
explosion which, on the other hand, increases the star angular momentum. According to \citet{watts}, the resulting period 
after one year or so, is largely independent on the period at birth and depends mainly on the neutron star magnetic field, 
with lower magnetic field and lower {\em r-modes} saturation amplitude resulting in lower periods (see section 4 of their paper and, in particular,
figures 2 and 4). They find that neutron stars with magnetic field $B\sim 10^{12}~G$ have final rotation periods, about one year after the supernova explosion, 
somewhere between $\sim 2~ms$ and $\sim 15~ms$. For GWDNS we could expect substantially lower values, 
but given all the uncertainties we prefer to take the period uniformly distributed in that range.   

\subsection{Age distribution}
\begin{figure}
\includegraphics[width=84mm]{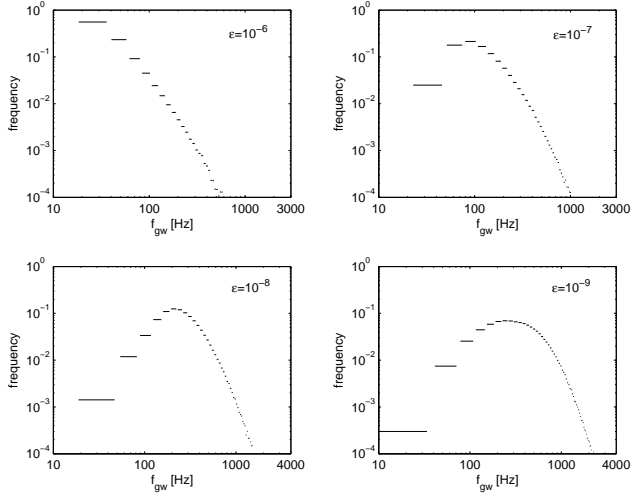}
%%\vspace{2cm}
\caption{Final emission frequency distribution for a population of GWDNS with ages up to $10~Gyr$, for different fixed values of the star ellipticity.\label{fgwfin}}
\end{figure}

The age of GWDNS is chosen from a uniform distribution between 0 and $100~Myr$ or $500~Myr$. 
The reason for these choices is the following.
We have computed the present frequency distribution of the signals emitted by a population of GWDNS, evolving according 
to Eq.(\ref{ftime}), for different values of the star ellipticity and assuming a maximum age $t_{{max}}=10~Gyr$. 
From Fig.(\ref{fgwfin}) we see that most of the sources emit in the detector frequency bandwidth, even for high values of 
the ellipticity. Then, in principle, we should simulate the population of GWDNS with ages up to the age of the Galaxy. 
In practice, however, we have verified that all detected sources in the simulation have ages much lower than that. 
In the case of the initial Virgo detector, in particular, most detectable sources have age less than about $1-2~Myr$. This
is a consequence of the fact that mainly fast rotating (and then young) objects can produce detectable signals. As already said
in Section 3, detectable GWDNS typically belong to the solar neighbourhood (see Section 5.1 for a more quantitative discussion).
A typical neutron star with a kick velocity of about $400~kms^{-1}$ will move, in that time, of about $600~pc$ respect 
to the Sun. Then, considering also the over production of neutron stars in the Gould Belt, sources today in the solar 
neighbourhood are likely to be born in the solar neighbourhood.   
To be conservative, we have chosen $t_{max}=100~Myr$ for the first generation of interferometric detectors and 
$t_{{max}}=500~Myr$ for advanced detectors. Assuming a constant neutron star gslsctic formation rate of 
$\sim 0.02~yr^{-1}$ we have that $N_{tot}\sim 2\cdot 10^6$ neutron stars have been formed in the last $100~Myr$ 
(or $10^7$ in $500~Myr$). The number of GWDNS will be a fraction $\lambda$ of $N_{{tot}}$ that we treat as a free parameter.
$N_{tot}$ is the number of neutron stars we have used in the simulations.  

\subsection{Ellipticity distribution}
The typical ellipticity of neutron stars is largely unknown and also the allowed maximum value, depending on the properties of matter at very high densities, is very uncertain. At present, maximum values $\epsilon_{max}\sim 10^{-7}\div 10^{-6}$ are considered plausible \citep{usho}. 
We cope with our poor knowledge by using, for the ellipticity, the probability distribution which satisfies the {\em principle of maximum entropy}, that is that leaves us with the largest uncertainty, without introducing any additional assumption or bias. The probability distribution $p(x)$ maximizing the differential entropy $H(x)=-\int p(x)\log{p(x)}dx$, under the constraint that the expectation value is given, is the exponential distribution
\begin{equation}
p(\epsilon)={e^{-{\epsilon \over{\tau}}}
\over{\tau \left(1-e^{-{\epsilon_{max}\over{\tau}}}\right)}}\label{elldist}  
\end{equation}
where $\epsilon_{max}$ is the maximum possible value for $\epsilon$. The relation between $\tau$ and the mean value of the distribution, $\bar{\epsilon}$, is 
\begin{equation}
\bar{\epsilon}=\tau-{\epsilon_{max}\over{e^{{\epsilon_{max}\over{\tau}}}-1}}
\end{equation}
In the simulation, chosen a value for the mean ellipticity $\overline{\epsilon}$, the parameter $\tau$ is calculated numerically using the previous equation. 
With the choice $\epsilon_{max}=2.5\cdot 10^{-6}$, the maximum mean value is $10^{-6}$. In Section 5, we will report results, concerning the detection
rate, for the different initial period distributions and for mean ellipticities $\overline{\epsilon}=10^{-6},10^{-7},10^{-8}$.

\subsection{Galactic gravitational potential energy}
We calculate the motion of GWDNS in the Galaxy using the following expressions for the galactic gravitational potential energy \citep{pacz}:
\begin{equation}
\Phi_i(R,z)=-{{GM_i}\over{\sqrt{R^2+[a_i+\sqrt{(z^2+b^2_i)}]^2}}}
\label{phii}
\end{equation}
where $i=1,2$ correspond respectively to the spheroid and the disk, and
\begin{equation}
\Phi_h(r)=-{{GM_c}\over r_c}\left(1+{1\over 2}\ln{(1+{r^2_h\over r^2_c})}-{1\over 2}\ln{(1+{r^2\over r^2_c})}-
{r_c\over r}\arctan{r\over r_c}\right)
\label{phih}
\end{equation}
which describes the galactic halo, with $r=\sqrt{x^2+y^2+z^2}$. Following \citet{gonth}, we have added the constant term in the halo potential, which does not affect star trajectories but is important if one wants to evaluate the number of objects which escape from the Galaxy. 
The parameters are
\begin{equation}
M_1=1.12\cdot 10^{10}~M_{\sun},~~a_1=0,~~b_1=0.277~kpc
\label{par1}
\end{equation}
\begin{equation}
M_2=8.07\cdot 10^{10}~M_{\sun},~~a_2=3.7,~~b_1=0.20~kpc
\label{par2}
\end{equation}
\begin{equation}
M_c=5.0\cdot 10^{10}~M_{\sun},~~r_c=6.0~kpc,~~r_h=41~kpc
\label{par3}
\end{equation}
From the galactic potential we calculate the rotation velocity as a function of the position $(R,z)$ (being the potential axisymmetric, the rotational curve does not depend on $\phi$) using the relation
\begin{equation}
v^2_{rot}=R{{\partial \Phi}\over{\partial R}}
\label{vrot}
\end{equation}
with $\Phi(R,z)=\Phi_1(R,z)+\Phi_2(R,z)+\Phi_h(R,z)$. We find 
\begin{equation}
v^2_{rot,i}=GM_i{R^2\over{\left(R^2+[a_i+\sqrt{z^2+b^2_i}]^2\right)^{3\over 2}}}
\label{vroti}
\end{equation}
\begin{equation}
v^2_{rot,h}={{GM_c}\over r_c}{R^2\over {R^2+z^2}}\left(1-{r_c\over \sqrt{R^2+z^2}}\arctan{ \left(\sqrt{R^2+z^2}\over r_c \right)}\right)
\label{vroth}
\end{equation}
The modulus of the total rotational velocity is $v_{rot}=\sqrt{v^2_{rot,1}+v^2_{rot,2}+v^2_{rot,h}}$.
Once we have extracted a triple $(v_{kick}, \phi_v, \theta_v)$ for the kick velocity and a triple $(R,z,\phi_0)$ for the star 
initial position, we integrate the equations of motion:
\begin{equation}
\ddot{R}=-{{\partial \Phi_{eff}}\over {\partial R}};~~
\ddot{z}=-{{\partial \Phi_{eff}}\over {\partial z}}
\label{eq1}
\end{equation}
with 
\[
\Phi_{eff}(R,z)=\Phi(R,z)+{L^2_z\over {2R^2}}
\]
being $L_z$ the z-component of the star angular momentum which is a constant of motion with value depending on $R,v_{rot},v_{kick},\phi_v,\phi_0,\theta_v$.
This system of two equations of the second order can be reduced to a system of four equations of the first order
which we numerically solve with initial conditions $R(t=0),~z(t=0),u(t=0)=\dot{R}(t=0),~v(t=0)=\dot{z}(t=0)$ using a fifth 
order Runge-Kutta method with adaptive step size. In the trajectory of each star the total energy is conserved to better 
than one part in $10^6$.

\section{Results of the simulations} \label{res}
We have done several simulations of a population of $N_{tot}=2\cdot 10^6$ GWDNS using the probability distributions 
previously described. $N_{tot}$ is the number of neutron stars we expect to be born in the Galaxy in the last $100~Myr$, 
assuming a constant formation rate of $0.02~yr^{-1}$, see the discussion in Section 4.4. In Table \ref{tab1} the properties of the different models are summarized. 
\begin{table}
%\begin{center}
\caption{Summary of model parameters. Concerning the initial period distribution function, $p(P_0)$, "standard" refers to the period distribution described by Eq.(\ref{p0dist}); "r-modes" corresponds to the same period distribution of Model 1 but in which all periods less than $10~ms$ are set to $10~ms$, while "r-modes+fall-back" describes a uniform distribution between $2~ms$ and $15~ms$, see Section \ref{initper} for a detailed description.\label{tab1}}
\begin{tabular}{ccc}
\hline
Model & $p(P_0)$ & $\overline{\epsilon}$ \\
\hline
$1$ & standard & $10^{-6}$ \\
$2$ & standard & $10^{-7}$ \\
$3$ & standard & $10^{-8}$ \\
$4$ & r-modes  & $10^{-6}$ \\
$5$ & r-modes  & $10^{-7}$ \\
$6$ & r-modes  & $10^{-8}$ \\
$7$ & r-modes+fall-back & $10^{-6}$ \\
$8$ & r-modes+fall-back & $10^{-7}$ \\
$9$ & r-modes+fall-back & $10^{-8}$ \\
\hline
\end{tabular}
%\end{center}
\end{table}
In each run the number of detected sources is computed, assuming the following search parameters, already discussed in Section \ref{detect}: total observation time $T_{obs}=4$ months, {\em full-sky} search, signal frequency up to $2~kHz$, minimum spin-down time $\tau=10^4~yr$, $10^9$ candidates selected after the first incoherent step. Some quantities related to the detected sourcers are also computed: the $50^{th}$ and $90^{th}$ percentiles of the signal frequency, distance, declination angle, and age distribution functions. 
To reduce statistical fluctuations results, for each model, are averaged among 300 independent runs for initial Virgo and over 50 independent runs for the advanced Virgo detector. Each run takes about 2 hours on a Xeon 2.4 processor machine. We have performed the whole set of simulations within the {\it Production Grid}\footnote{http://grid-it.cnaf.infn.it/} of the italian {\it Istituto Nazionale di Fisica Nucleare}, consisting of a large set of geographically distributed computational and storage resources covering more than 20 sites. 
This has largely reduced the time needed for the computations. 

Independently of the model, we have found that $\sim 45\%$ of the neutron stars population will escape from the Galaxy, 
due to the high kick velocity. This is a well-known result found, e.g., by \citet{arzo}. 

According to \citet{popov2}, the number of neutron stars within $\sim 1~kpc$ from the Sun and with age less than $\sim 4~Myr$ should be, at present, around $100$. The number we find is $\sim 140$, comparable with that. Of these, $\sim {1\over 2}$ were born in the Gould Belt. Most of the detected sources, with the first generation of interferometers, have ages less than that value. Then, the presence of the Gould Belt impacts in a significant way on the expected number of detections. 

A negligible fraction of the detected sources has a spin-down decay time less than $10^4~yr$.

Let us now discuss in more detail the results of the simulations considering separately 
first and second generation detectors. We use the sensitivity curves of 
Virgo detector and of a possible advanced Virgo detector, shown in Figs.(\ref{pulsar1},\ref{pulsar2}).

\subsection{Results for first generation interferometers}
Results are summarized in Tab.(\ref{tab2}). The number of detected sources is parametrized by the fraction $\lambda \le 1$ of the total neutron star population made of GWDNS. Only the models giving $N_{det}\ga 1\cdot \lambda$ are shown. For Model 1, for instance, we see that assuming $\lambda=0.1$, i.e. 
that $10\%$ of all neutron stars are GWDNS, we would expect to have $\sim 1$ detection. More precisely, if Model 1 holds, 
the minimum fraction $\lambda_{min}$ we need to have at least one detectable source is $7.6\%$.  
\begin{table*}
%\begin{center}
\begin{minipage}{140mm}
\caption{Results for Virgo detector. The first column indicates the model, according to the classification of Tab.(\ref{tab1}); the second column shows the number of expected detections in units of $\lambda$, which is the fraction of the total neutron star population made of GWDNS; $f_{50},r_{50},|\theta|_{50},t_{50}$ and $f_{90},r_{90},|\theta|_{90},t_{90}$ give the values of the gravitational wave frequency, of the distance from the Sun, of the modulus of the declination and of the age such that respectively $50\%$ and $90\%$ of the detected sources have values lower than those. Only models giving $N_{det}\ga 1\cdot \lambda$ are shown.\label{tab2}}
\begin{tabular}{cclllll}
\hline
Model & ${N_{det}\over \lambda}$ & $f_{50},~~f_{90}[Hz]$ & $r_{50},~~~r_{90}[kpc]$ & $|\theta|_{50},~~~|\theta|_{90}[deg]$ & $t_{50},~~~t_{90}[Myr]$ & \\
\hline
$1$ & $13.2$ & $288,~473$ & $0.25,~0.45$ & $18.3,~49.8$ & $0.33,~1.15$ \\
$2$ & $0.84$ & $490,~954$ & $0.14,~0.25$ & $18.3,~54.4$ & $0.43,~2.00$ \\
$4$ & $3.4$ & $190,~198$ & $0.13,~0.23$ & $20.6,~53.8$ & $0.25,~1.48$  \\
$7$ & $8.3$ & $255,~425$ & $0.19,~0.41$ & $18.9,~52.7$ & $0.34,~1.43$  \\
$8$ & $0.82$& $452,~698$ & $0.13,~0.26$ & $22.9,~61.9$ & $0.53,~2.05$  \\
\hline
\end{tabular}
%\end{center}
\end{minipage}
\end{table*}
\begin{figure}
\includegraphics[width=84mm]{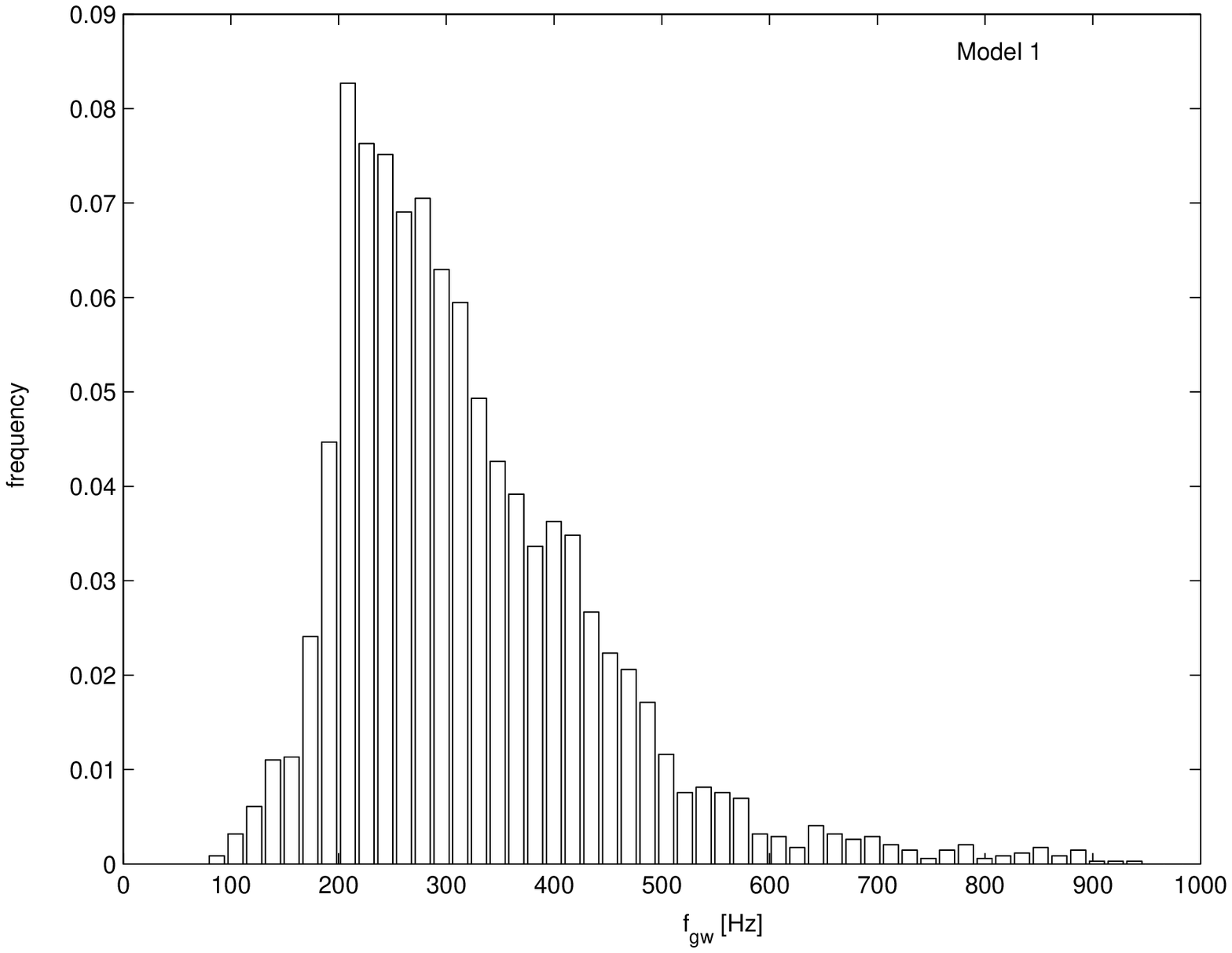}\includegraphics[width=84mm]{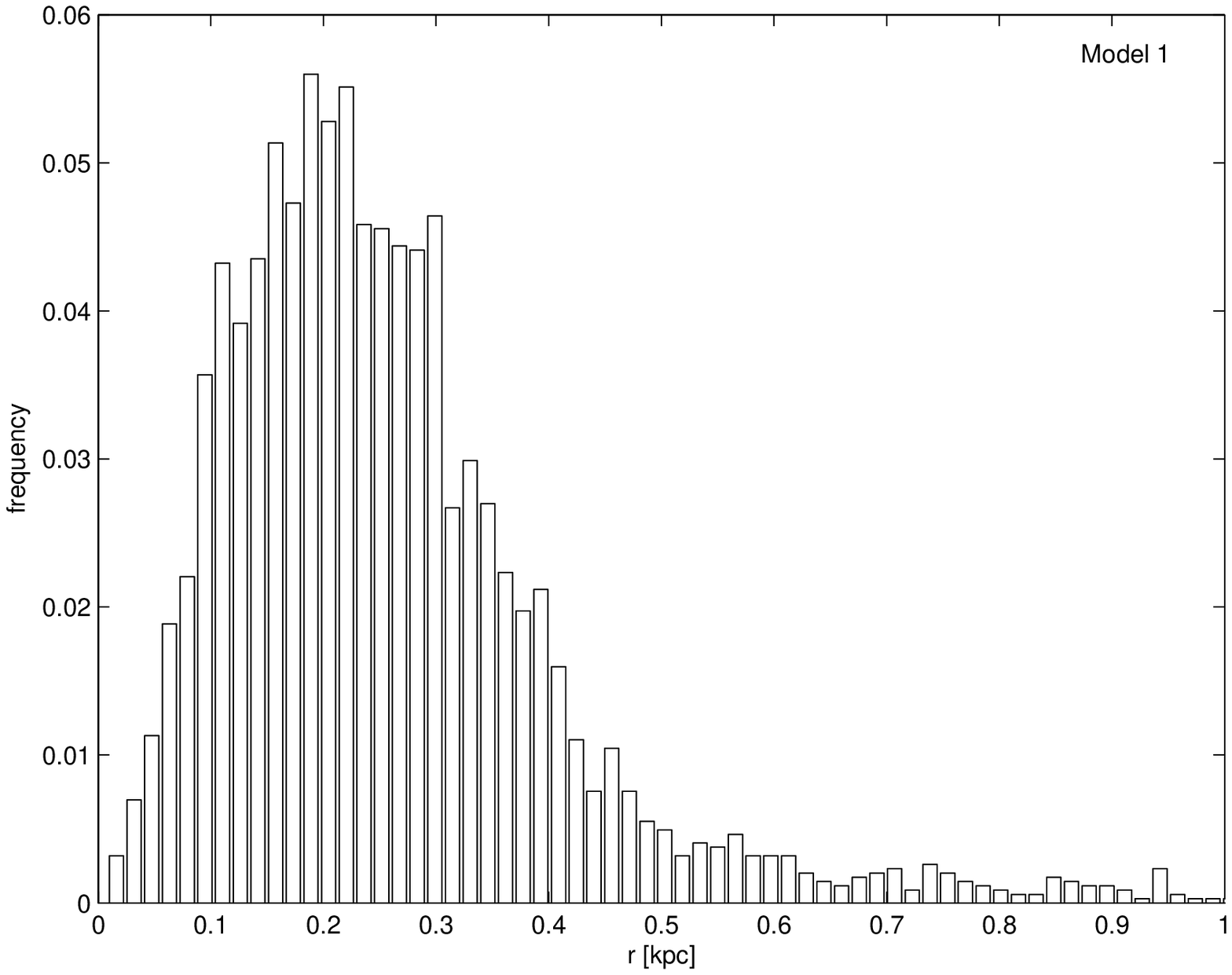}
%%\vspace{2cm}
\caption{Model 1: distribution of the gravitational wave frequency and distance from the Sun for detected sources with the Virgo detector.\label{fgw_6_std}}
\end{figure}
\begin{figure}
\includegraphics[width=84mm]{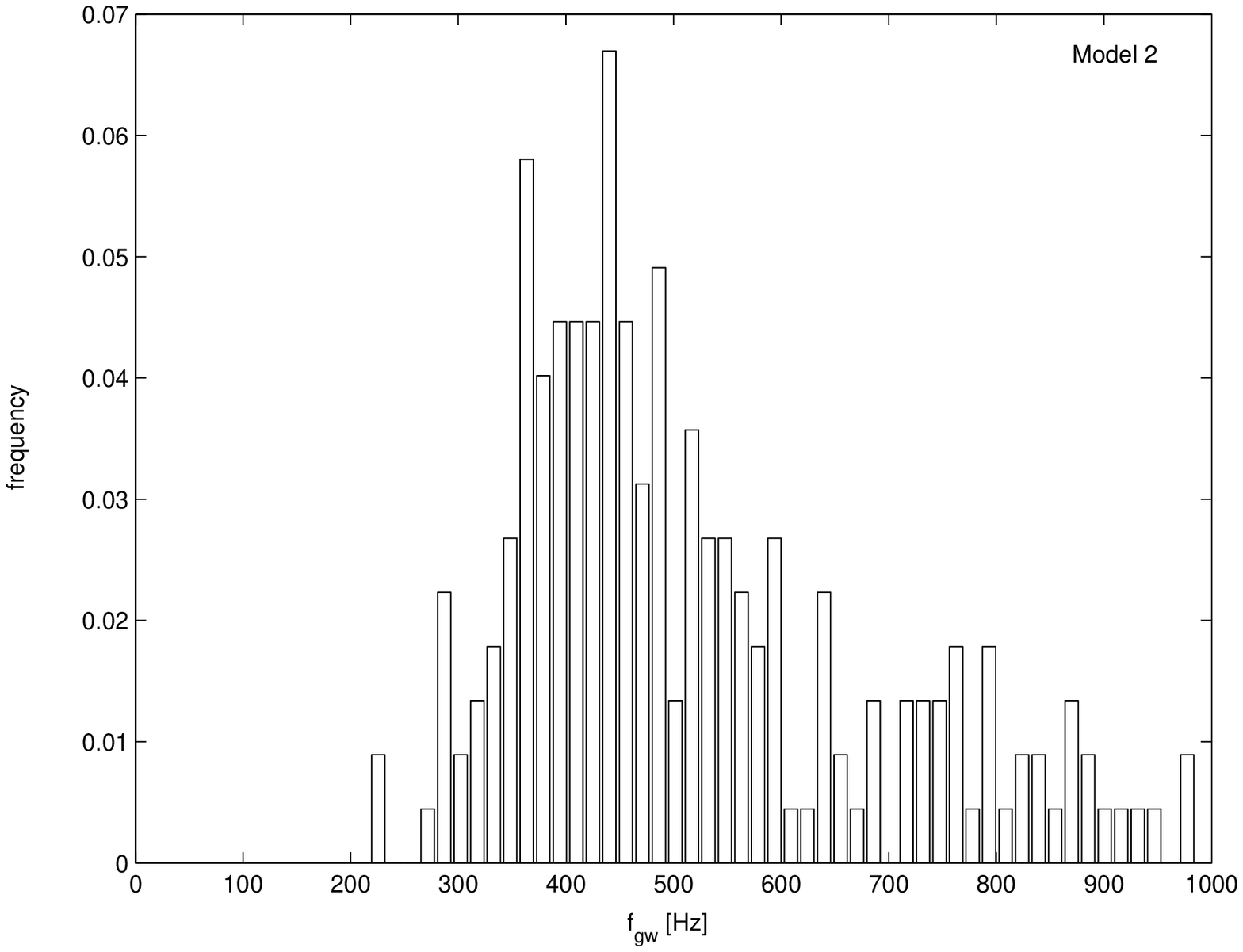}\includegraphics[width=84mm]{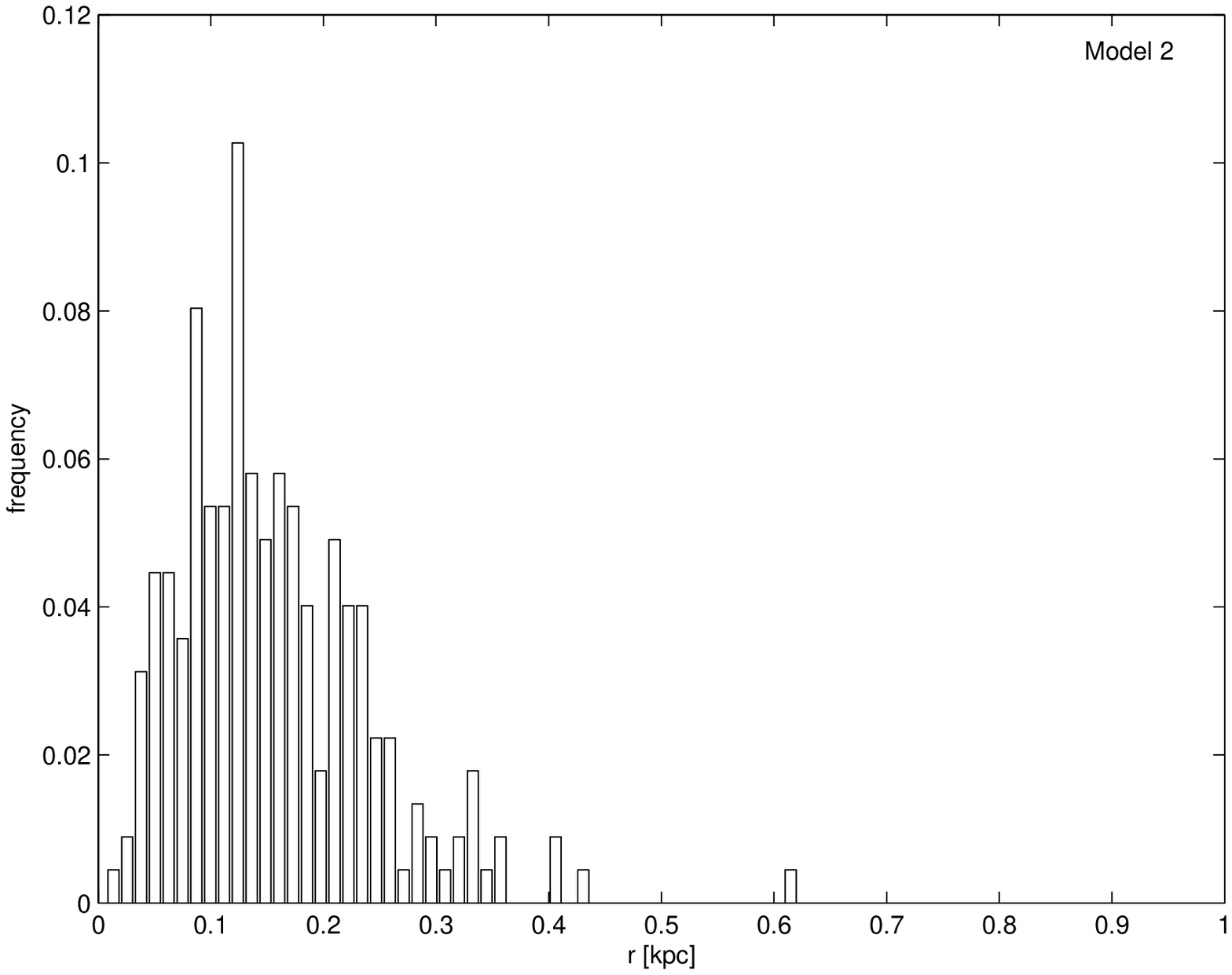}
%%\vspace{2cm}
\caption{Model 2: distribution of the gravitational wave frequency and distance from the Sun for detected sources with the Virgo detector. \label{fgw_7_std}}
\end{figure}
On the other hand, for Model 2 we would expect $\sim 1$ detection if {\it all} the neutron 
star population were made of GWDNS, i.e. $\lambda=1$, which could happen only if there were some mechanism suppressing the 
external magnetic field of a young neutron star, see Section \ref{lowb}. Typically, detected sources rotate at high rate and are very near ($f_{gw}>200~Hz$ and $r<500~pc$ for most sources). Most detectable sources, moreover, are very young, with age less than $\sim 2~Myr$ and then, some of them could be detectable by ROSAT through their thermal emission, see e.g. \citet{pop}. In such a short time interval the majority of GWDNS have moved for a rather short distance from their initial position, 
few hundreds of Megaparsec at the most, and then many detected objects were born in the solar neighbourhood and still 
belong to the local disk or to the Gould Belt. This explains why for all models about $50\%$ of the detected sources have 
declination, in modulus, $\la 18^o$, which is the inclination of the Gould Belt, see Section \ref{spa}. 
If we consider only the loudest sources, however, the spread in declination is larger: half of the $1\%$ loudest sources 
have a declination, in modulus, larger than $\sim 26^o$. In Fig.(\ref{fgw_6_std}) the gravitational wave frequency and 
distance distributions of detected GWDNS are plotted for Model 1; in Fig.(\ref{fgw_7_std}) the same quantities are plotted for Model 2. 
For a given period distribution, the smaller is the mean ellipticity and the larger are the typical rotation frequencies 
of detected sources and the smaller are the distances. For Model 1, for instance, the minimum gravitational wave frequency is $\sim 80~Hz$ while it is larger than $200~Hz$ for Model 2. Similarly, for Model 1 we have detected sources up to a distance from the Sun of about $8~kpc$, even if, as already said, almost all are much nearer; for Model 2 the maximum distance is less than $0.7~kpc$. No detection is obtained for Model 3.
\begin{figure}
\includegraphics[width=84mm]{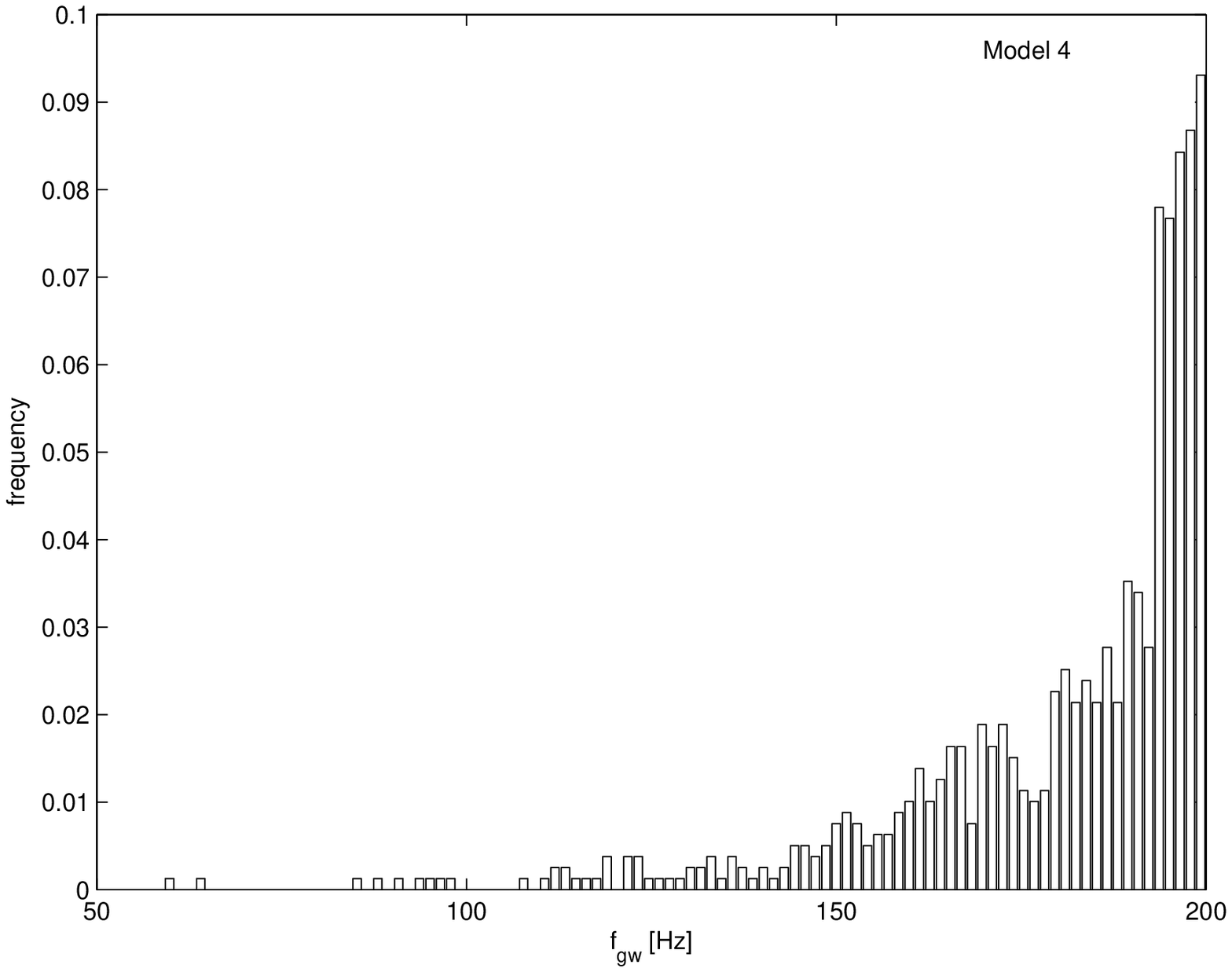}\includegraphics[width=84mm]{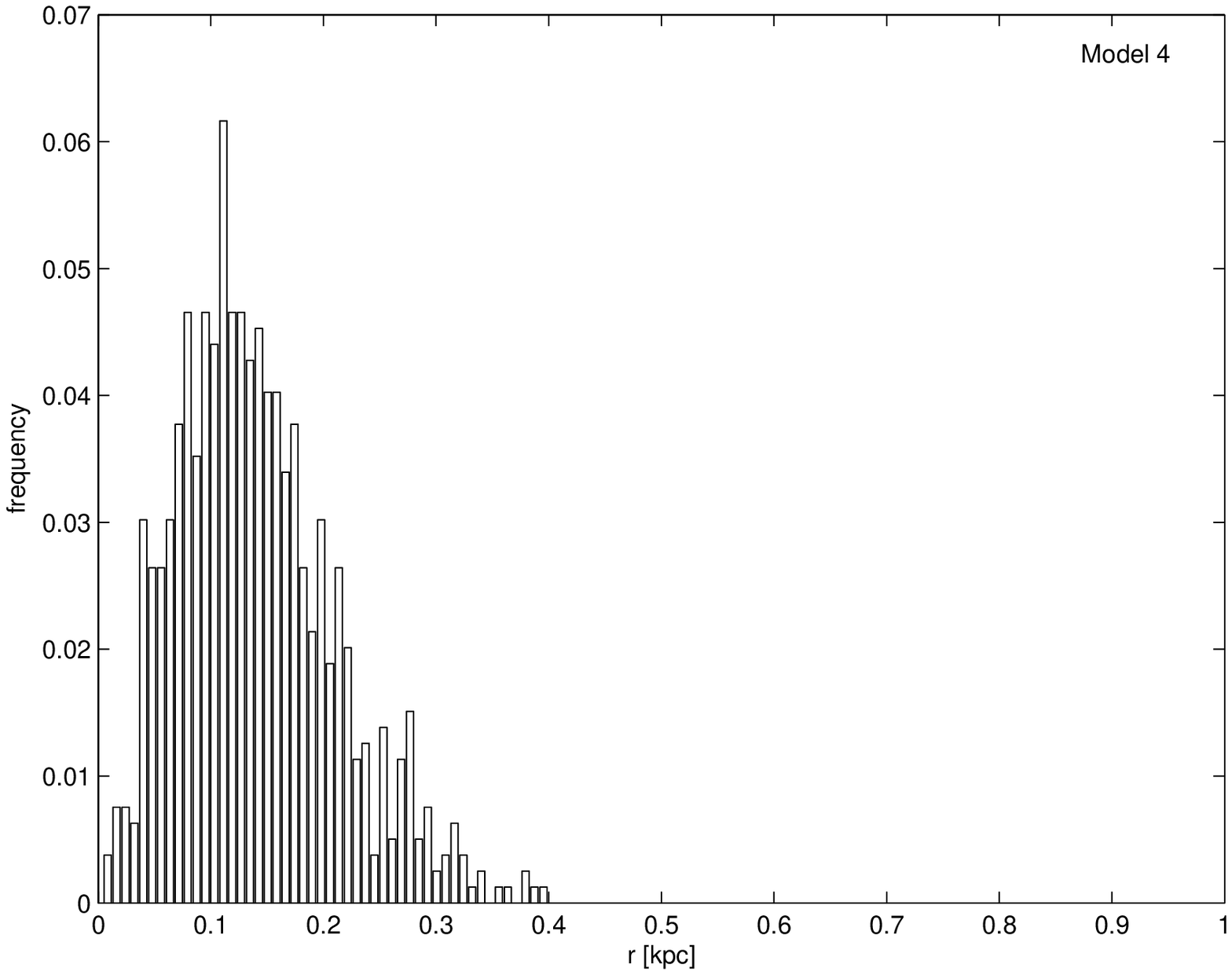}
%%\vspace{2cm}
\caption{Model 4: Distribution of the gravitational wave frequency and distance from the Sun for detected sources with the Virgo detector. \label{fgw_6_r}}
\end{figure}
\begin{figure}
\includegraphics[width=84mm]{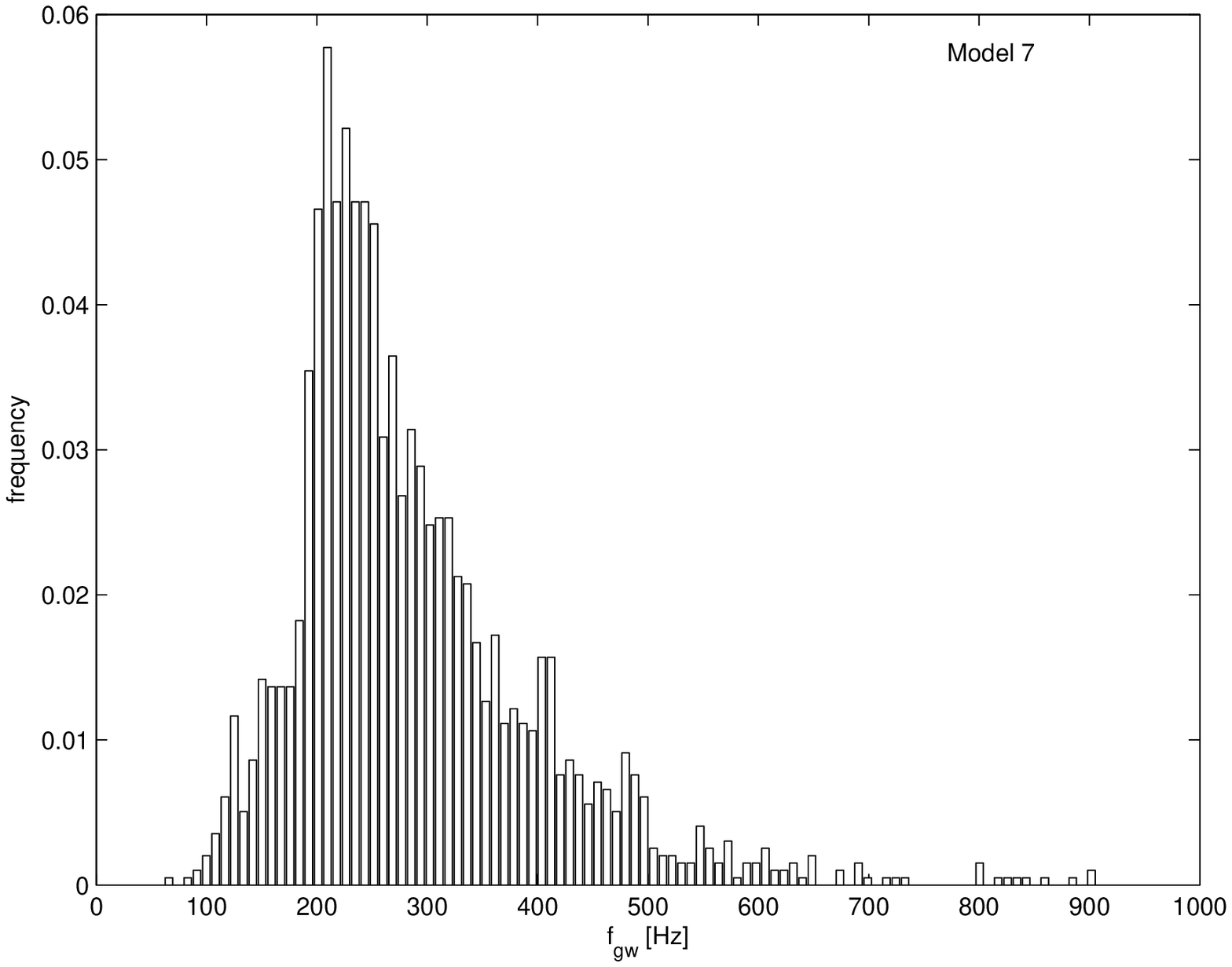}\includegraphics[width=84mm]{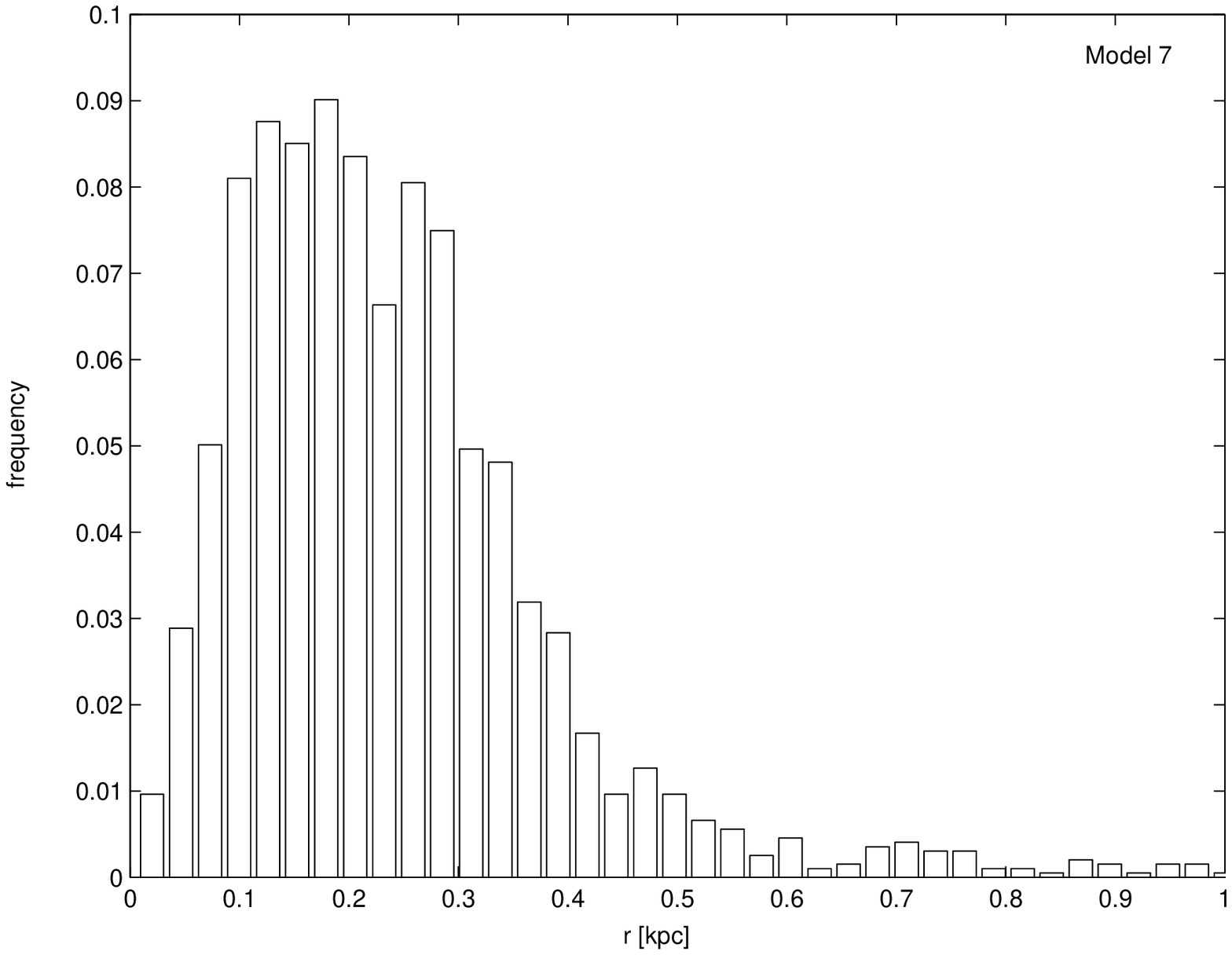}
%%\vspace{2cm}
\caption{Model 7: distribution of the gravitational wave frequency and distance from the Sun for detected sources with the Virgo detector. \label{fgw_6_unif}}
\end{figure}
\begin{figure}
\includegraphics[width=84mm]{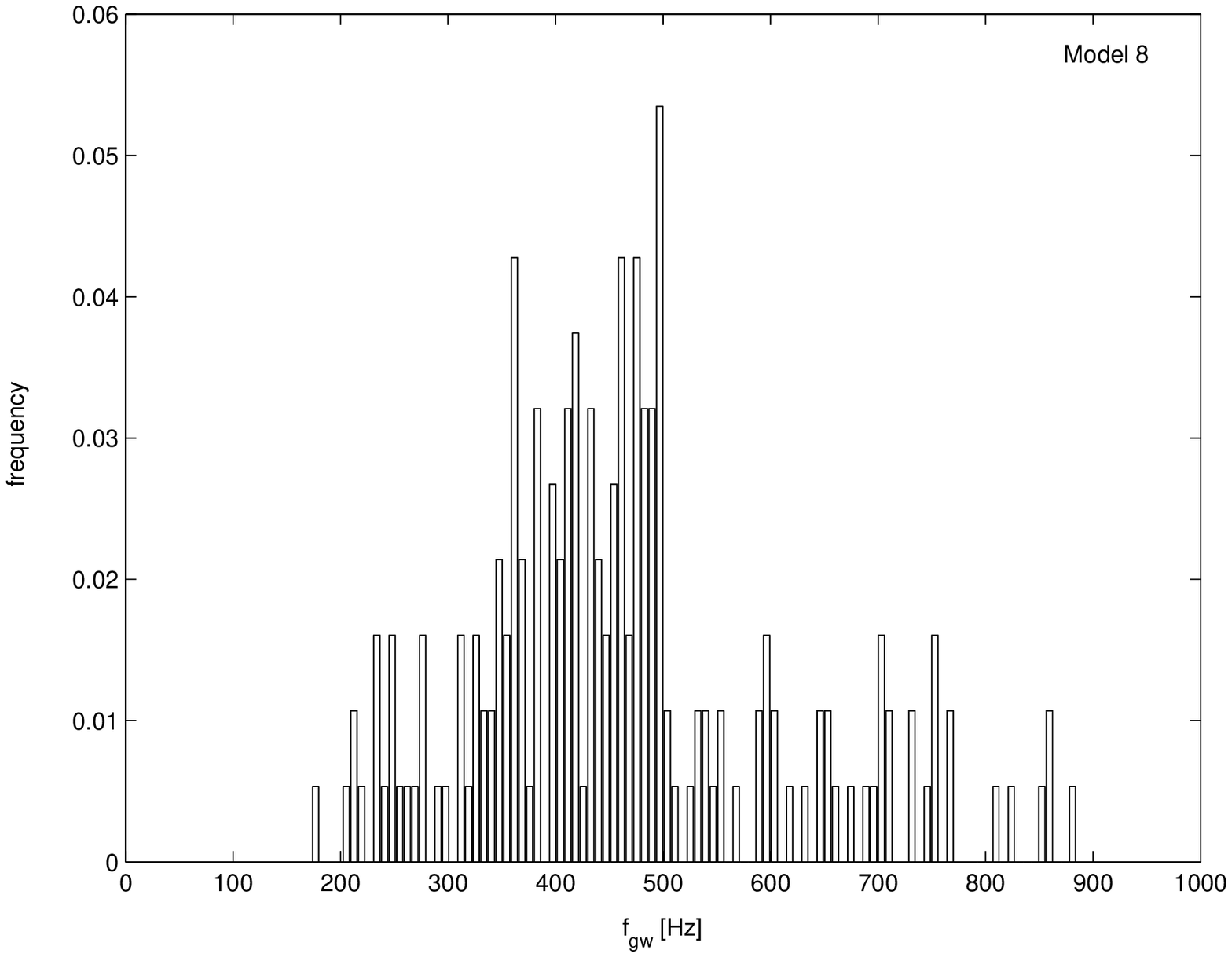}\includegraphics[width=84mm]{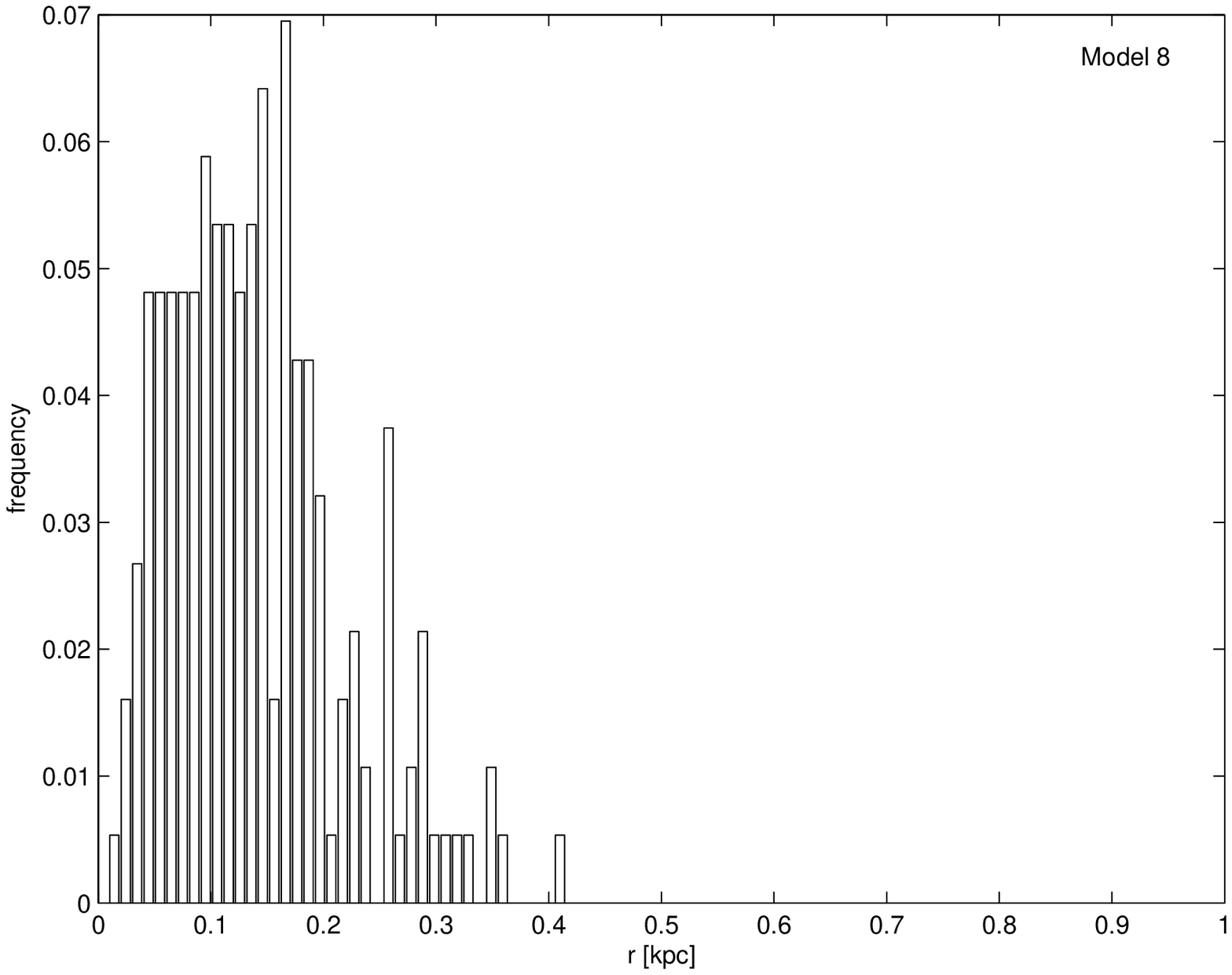}
%%\vspace{2cm}
\caption{Model 8: distribution of the gravitational wave frequency and distance from the Sun for detected sources with the Virgo detector. \label{fgw_7_unif}}
\end{figure}
Fig.(\ref{fgw_6_r}) refers to Model 4: note that the frequencies accumulate near $200~Hz$ which is the double of the maximum rotation frequency for that model; as the frequencies are on average lower than in Model 1, even if the mean ellipticity is the same, detected sources are on average nearer to the Sun. The minimum fraction we need in order to have at least one detected source is 
$\lambda_{min}=29.4\%$. Models 5 and 6 do not give any detection. 
Perspective of detection for this kind of initial period distribution, would be obvioulsy even worse if the cut-off 
initial period were larger than $10~ms$.
Fig.(\ref{fgw_6_unif}) refers to Model 7. The frequency distribution is similar to that of Model 1, but a little narrower 
and no source is detected at frequencies larger than $\sim 1~kHz$. 
Also the distance distribution is similar, but narrower, than that of Model 1. The minimum fraction $\lambda_{min}$
we would need to have -at least one detectable source is $12\%$.
Fig.(\ref{fgw_7_unif}) refers to Model 8: from the figure and Tab.(\ref{tab2}) we see that, as expected, detected sources 
are very near and emits signals at very high frequency. Like for Model 2, we could expect, at the most, 
one detection if {\em all} neutron stars were GWDNS. Model 9 produces no detection. In Tab.(\ref{tabeps}) we have reported, 
for each initial period distribution function and for three different values of the fraction $\lambda$, the minimum value 
of the neutron star mean ellipticity needed in order to 
have at least one detectable source; the minimum needed fraction $\lambda_{min}$, assuming $\overline{\epsilon}=10^{-6}$, 
is also shown.
\begin{table*}
%\begin{center}
\begin{minipage}{100mm}
\caption{For each kind of initial period distribution and for three different values of the fraction $\lambda$, 
in columns $2-4$ the minimum allowed value of the mean ellipticity 
$\overline{\epsilon}$ in order to have at least one detectable source is reported; in column $5$ there is the minimum 
allowed fraction if $\overline{\epsilon}=10^{-6}$. The target Virgo sensitivity is assumed.\label{tabeps}}
\begin{tabular}{ccccc}
\hline
$p(P_0)$ & $\lambda=0.5$ & $\lambda=0.2$ & $\lambda=0.1$ & $\lambda_{min}$ \\
\hline
standard & $1.4\cdot 10^{-7}$ & $2.5\cdot 10^{-7}$ & $4.8\cdot 10^{-7}$ & $0.076$ \\
r-modes &  $6.4\cdot 10^{-7}$ & $-$ & $-$ & $0.27$ \\
r-modes+fall-back & $2.4\cdot 10^{-7}$ & $6.0\cdot 10^{-7}$ & $-$ & $0.12$ \\
\hline
\end{tabular}
%\end{center}
\end{minipage}
\end{table*}
Only for $\lambda=0.5$ all the considered initial period distributions give at least one detection, with mean ellipticity 
in the range $\sim 1\div 6\cdot 10^{-7}$. For $\lambda=0.1$ only the "standard" period distribution gives at least
one detectable source 
(with $\epsilon \sim 5\cdot 10^{-7}$). Assuming $\overline{\epsilon}=10^{-6}$, the minimum needed fraction varies 
between $7.6\%$ to $27\%$.

\subsection{Results for second generation interferometers}
\begin{table*}
%\begin{center}
\begin{minipage}{140mm}
\caption{Results for an advanced Virgo detector. The first column indicates the model, according to the classification of Tab.(\ref{tab1}); $f_{50},r_{50},|\theta|_{50},t_{50}$ and $f_{90},r_{90},|\theta|_{90},t_{90}$ give the values of the gravitational wave frequency, of the distance from the Sun, of the modules of the declination and of the age such that respectively $50\%$ and $90\%$ of the detected sources have values lower than those. The parameter $\lambda$ is the fraction of the total neutron star population made of GWDNS. Only models giving $N_{det}\ga 1\cdot \lambda$ are shown.\label{tab4}}
\begin{tabular}{cclllll}
\hline
Model & ${N_{det}\over \lambda}$ & $f_{50},~~f_{90}[Hz]$ & $r_{50},~~~r_{90}[kpc]$ & $\theta_{50},~~~\theta_{90}[deg]$ & $t_{50},~~~t_{90}[Myr]$ \\
\hline
$1$ & $1174$ & $295,~470$ & $3.8,~8.7$ & $1.03,~34.4$ & $0.17,~5.70$  \\
$2$ & $169$ & $378,~638$ & $0.51,~1.7$ & $19.7,~60.2$ & $3.65,~54.3$ \\
$3$ & $9.4$ & $518,~1245$ & $0.14,~0.36$ & $25.2,~66.2$ & $17.0,~314$  \\
$4$ & $240$ & $149,~195$ & $0.56,~1.35$ & $20.1,~60.2$ & $2.8,~33.4$ \\
$5$ & $19.5$ & $196,~199$ & $0.2,~0.44$ & $22.6,~59.0$ & $2.4,~189$  \\
$7$ & $635$ & $268,~449$ & $2.26,~8.5$ & $2.3,~43.8$ & $.38,~9.2$  \\
$8$ & $82.2$ & $334,~525$ & $0.45,~1.4$ & $22.6,~58.3$ & $4.3,~80$  \\
$9$ & $3.6$ & $440,~812$ & $0.14,~0.33$ & $25.4,~63.9$ & $51.3,~399$  \\
\hline
\end{tabular}
%\end{center}
\end{minipage}
\end{table*}
\begin{figure}
\includegraphics[width=84mm]{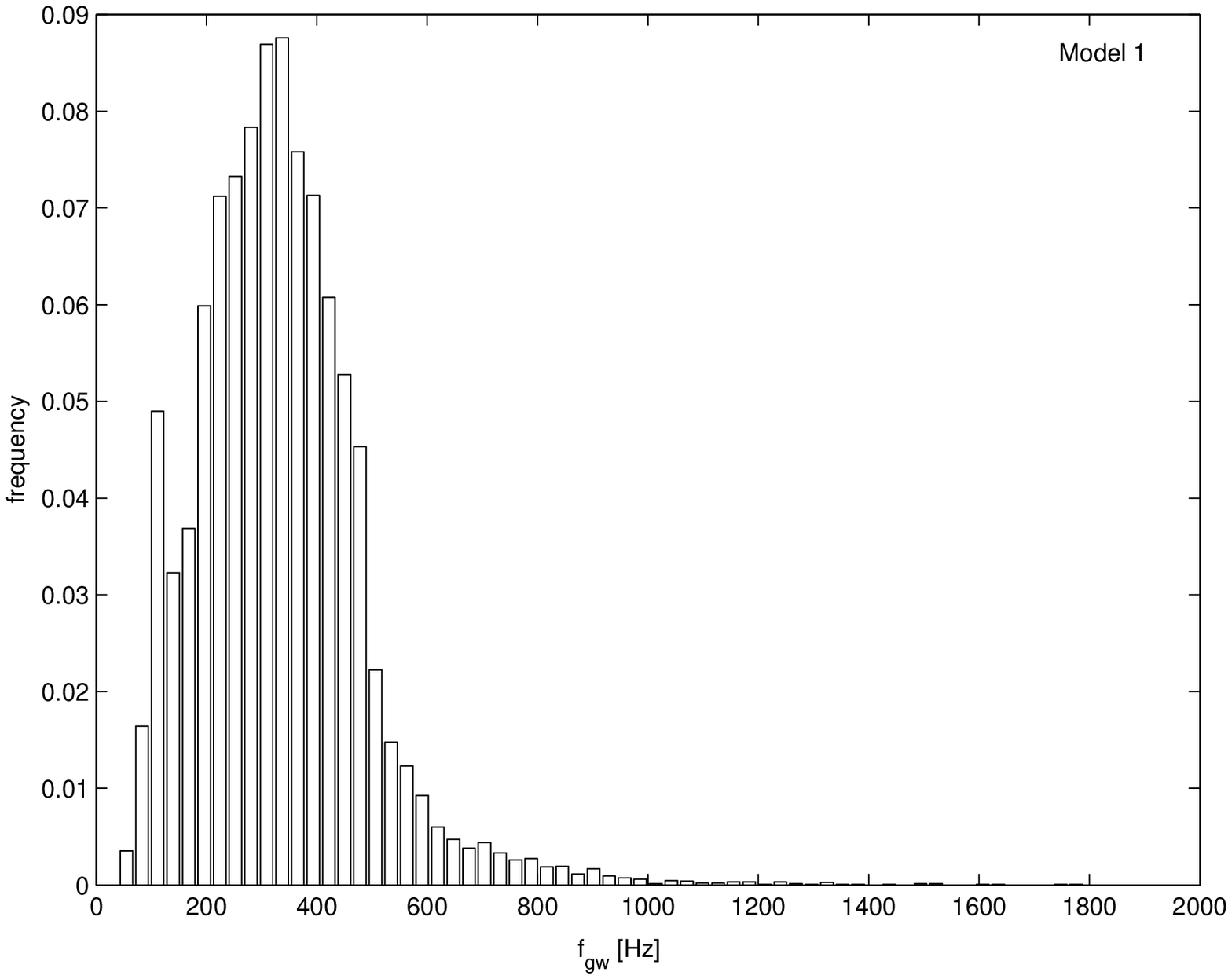}\includegraphics[width=84mm]{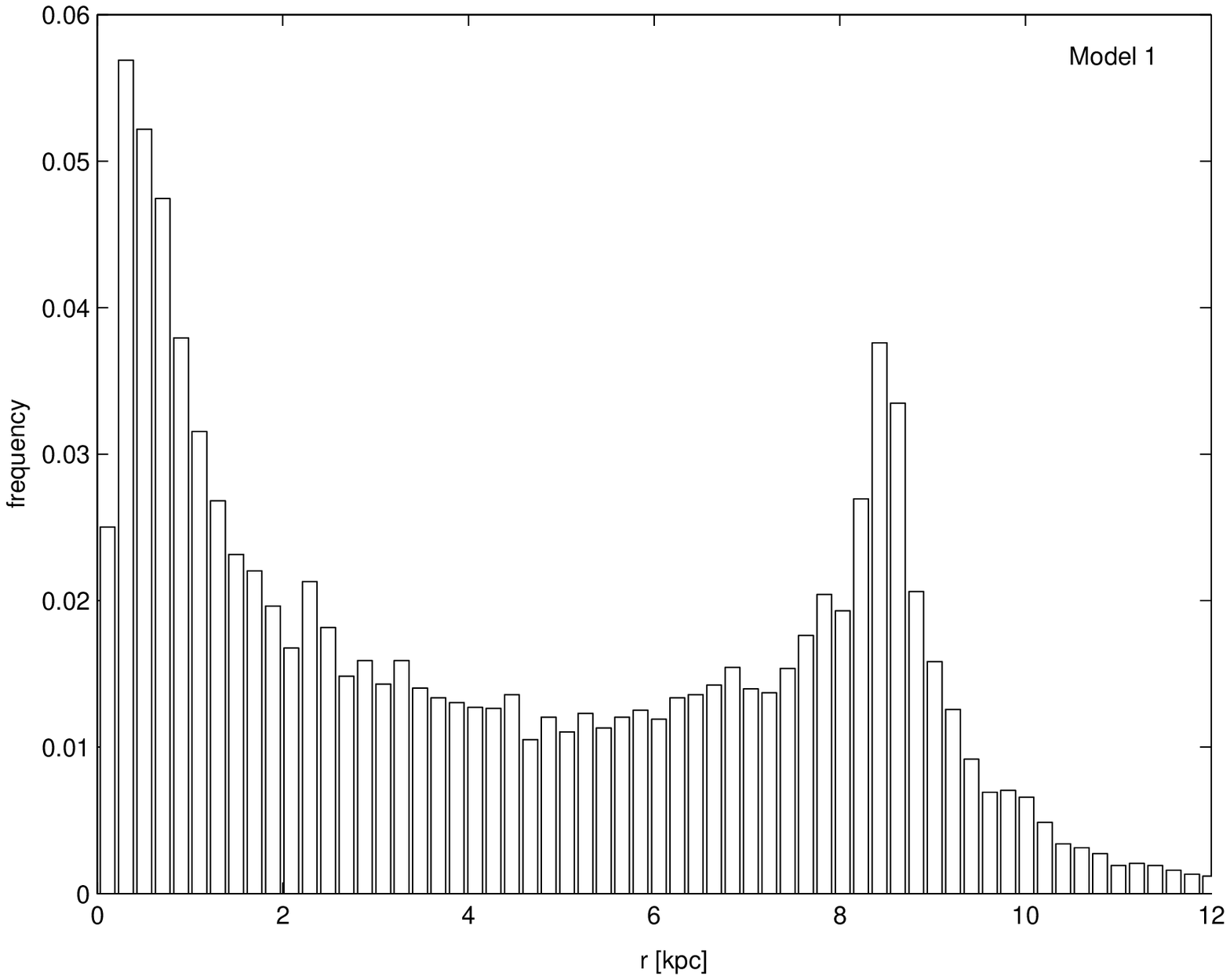}
%%\vspace{2cm}
\caption{Model 1: distribution of the gravitational wave frequency and distance from the Sun for detected sources with the advanced Virgo detector. \label{fgw_6_std_adv}}
\end{figure}
\begin{figure}
\includegraphics[width=84mm]{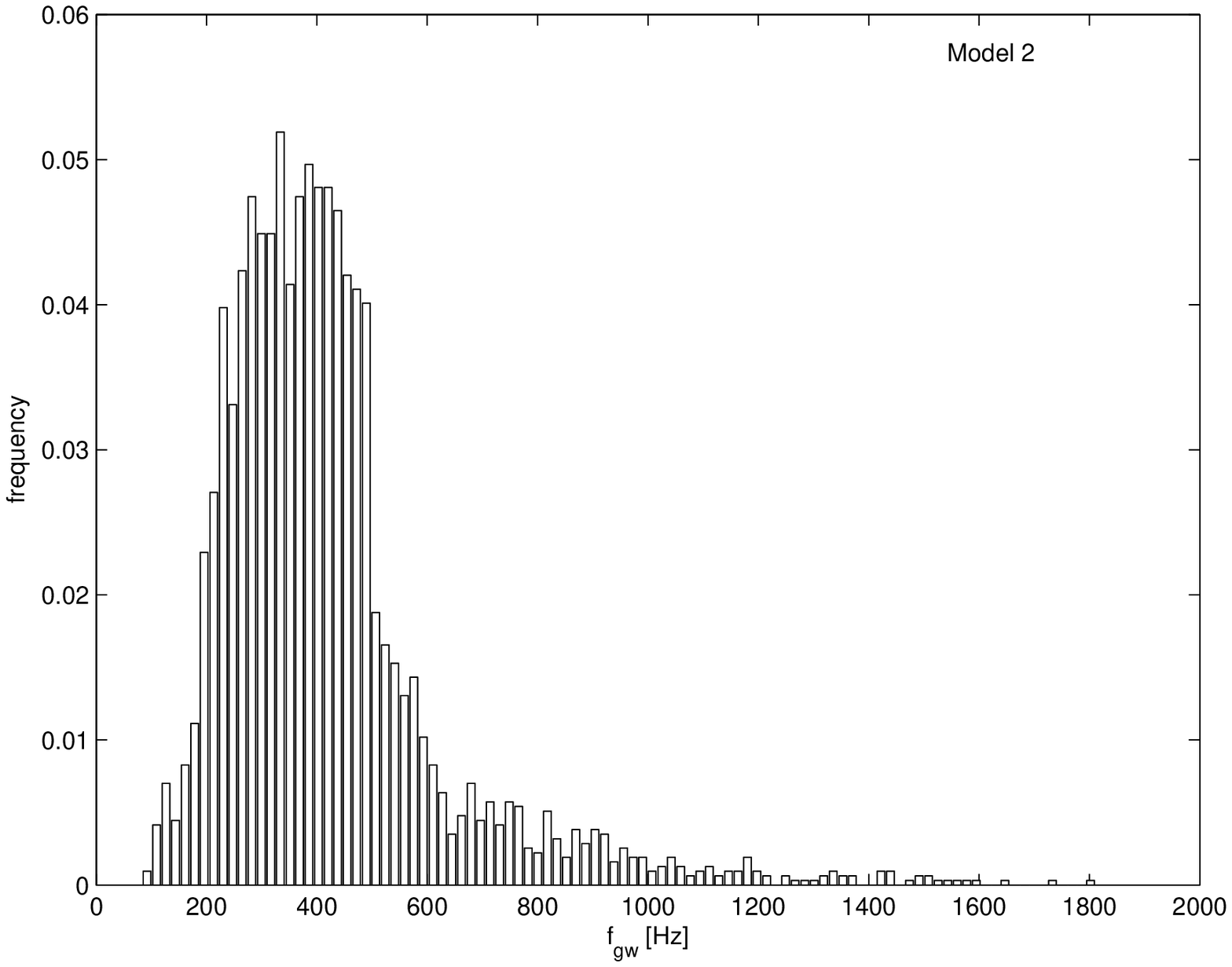}\includegraphics[width=84mm]{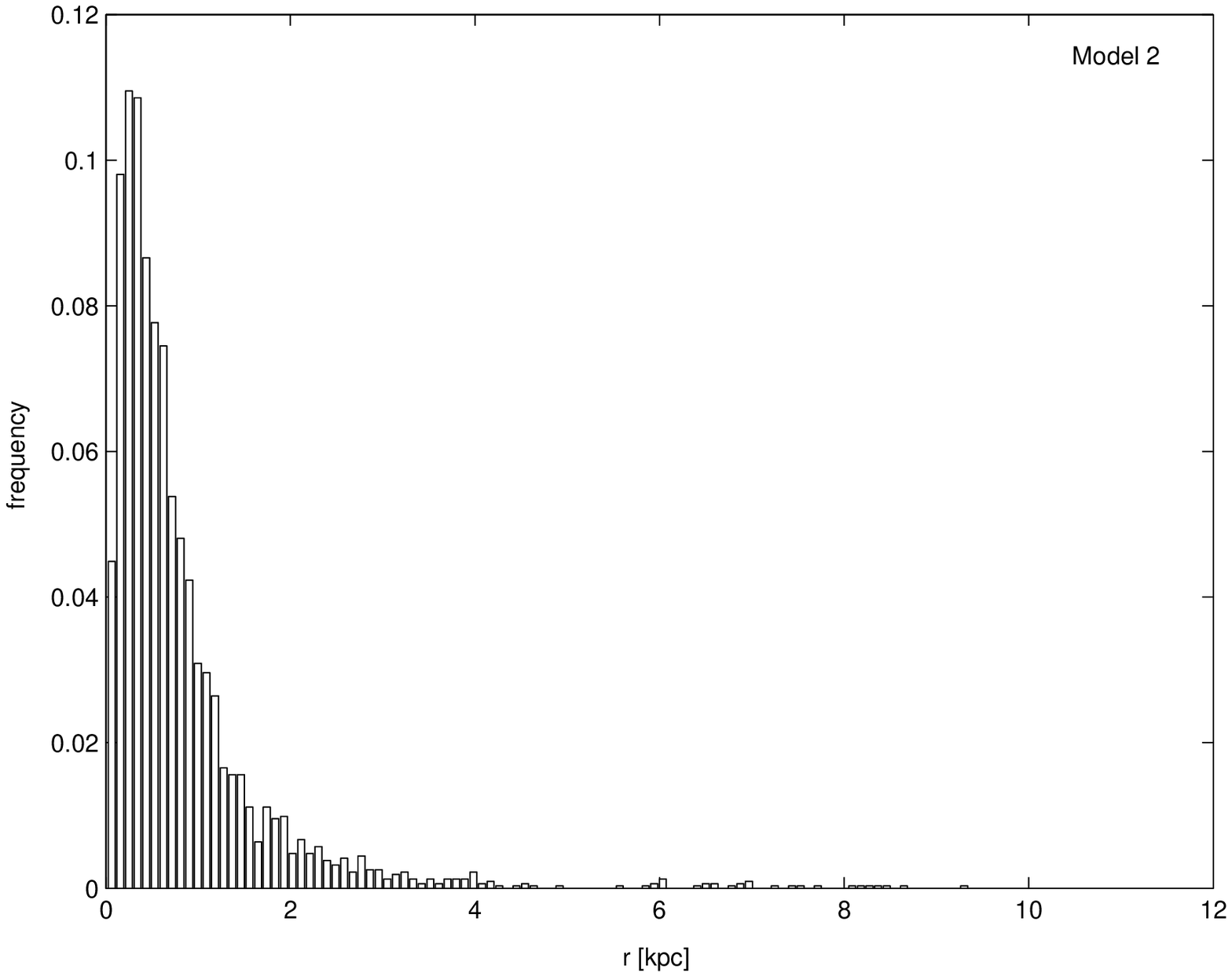}
%%\vspace{2cm}
\caption{Model 2: distribution of the gravitational wave frequency and distance from the Sun for detected sources with the advanced Virgo detector. \label{fgw_7_std_adv}}
\end{figure}
\begin{figure}
\includegraphics[width=84mm]{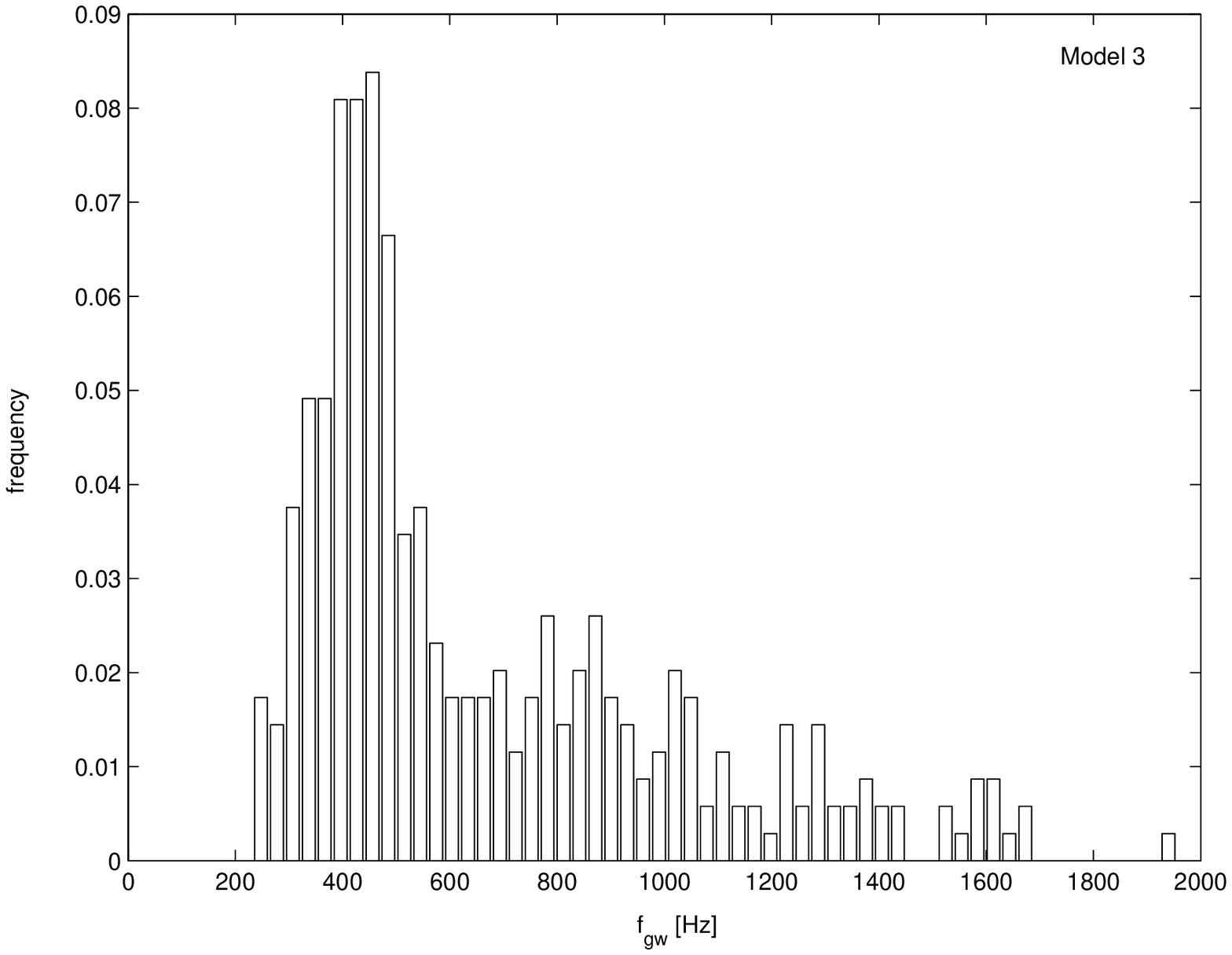}\includegraphics[width=84mm]{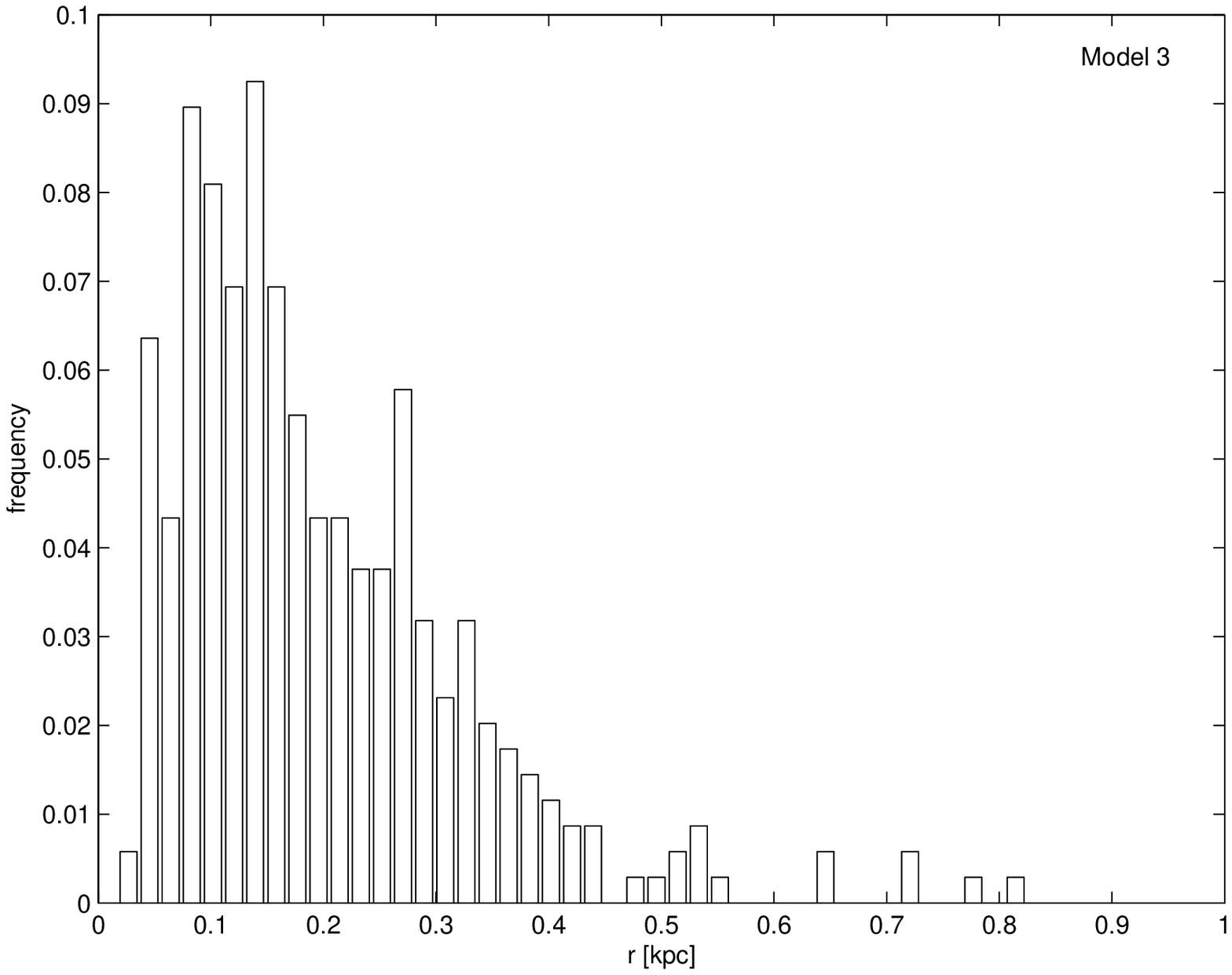}
%%\vspace{2cm}
\caption{Model 3: distribution of the gravitational wave frequency and distance from the Sun for detected sources with the advanced Virgo detector. \label{fgw_8_std_adv}}
\end{figure}
Results are summarized in Tab.(\ref{tab4}). The detection rate increases by about two orders of magnitude respect to the case of detectors of the first generation.
Fig.(\ref{fgw_6_std_adv}) refers to Model 1. From it and from Tab.(\ref{tab4}) we see that, while the frequency distribution is not heavily different from that concerning Model 1 for the Virgo detector, strong differences are evident for the other parameters. In particular, the distance distribution shows a large peak at small distances but also a second peak, with amplitude less than $50\%$ lower, around $8.5~kpc$, corresponding to the galactic centre. In any case, from the values of $r_{50}$ and $r_{90}$ it is clear that most of detected sources belong, in this case, to the galactic disk, and the contribution of the Gould 
Belt is small. This conclusion is reinforced by the value of $|\theta|_{50}$, which is around $1^o$. However, if the 
loudest sources are taken into account, half of the $1\%$ strongest emitters have declination larger, in modulus, than 
$\sim 16.7^o$, comparable with the values we obtain for the Virgo detector, and belong to the local neighbourhood. 
Another difference, respect to the case of the Virgo detector, is that now we have a non negligible fraction of detected GWDNS with ages larger than $\sim 5~Myr$: this is an obvious consequence of the fact that due to the higher sensitivity we can detect sources emitting at lower frequencies, and then older.   
Fig.(\ref{fgw_7_std_adv}) and Fig.(\ref{fgw_8_std_adv}) refer, respectively, to Model 2 and Model 3. As expected, the 
characteristic frequencies of the detected sources increase and the typical distances decrease, in order to compensate 
the smaller ellipticity. While for Model 2 we have 
still $\sim 10\%$ of the detected sources at $r\ga 2~kpc$, for Model 3 almost all the sources are in the solar 
neighbourhood. Another important difference respect to Model 1, and especially respect to the corresponding models for 
detectors of the first generation, is the characteristic age of detected objects which is now of the order of $100~Myr$ or 
even more. 
\begin{figure}
\includegraphics[width=84mm]{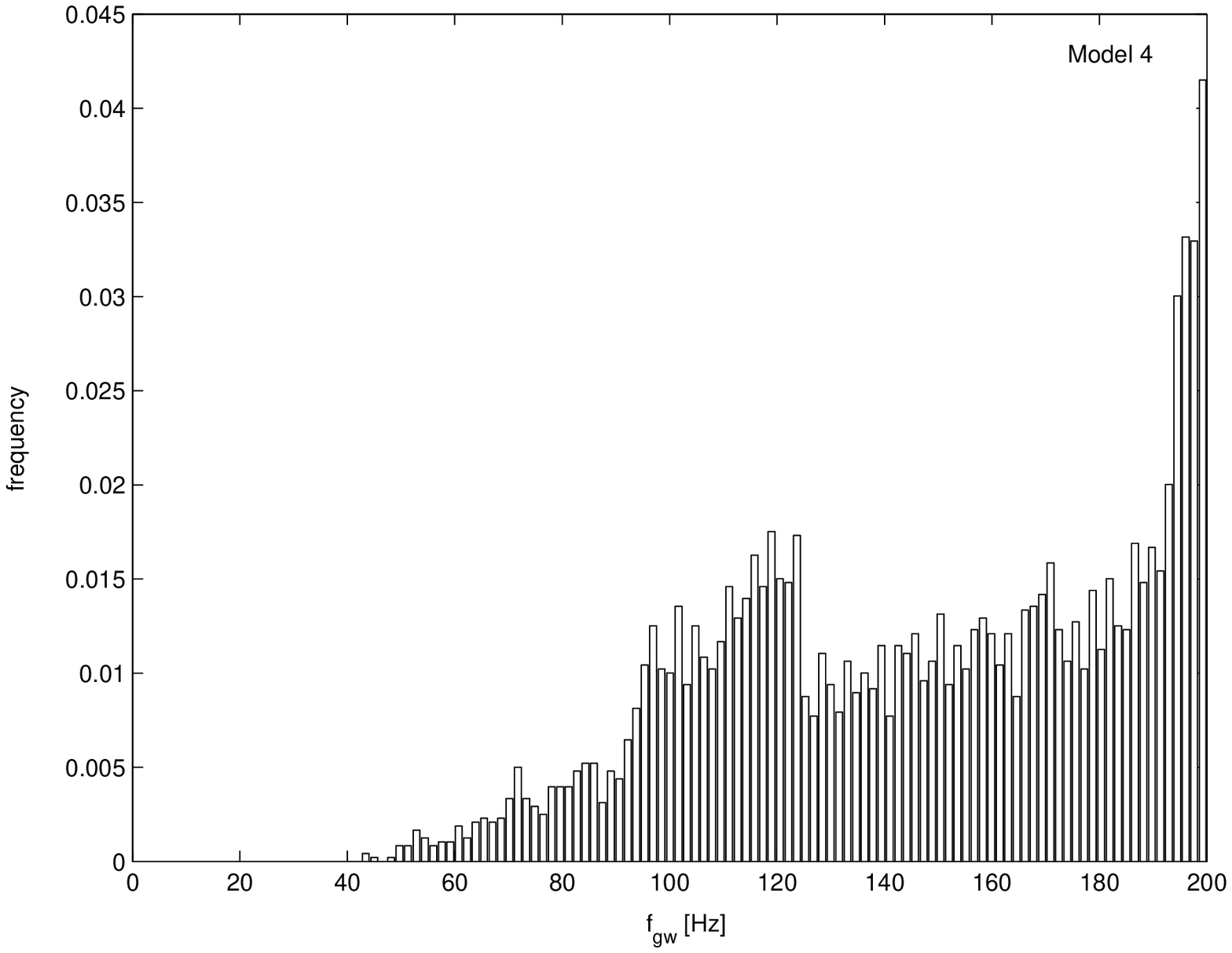}\includegraphics[width=84mm]{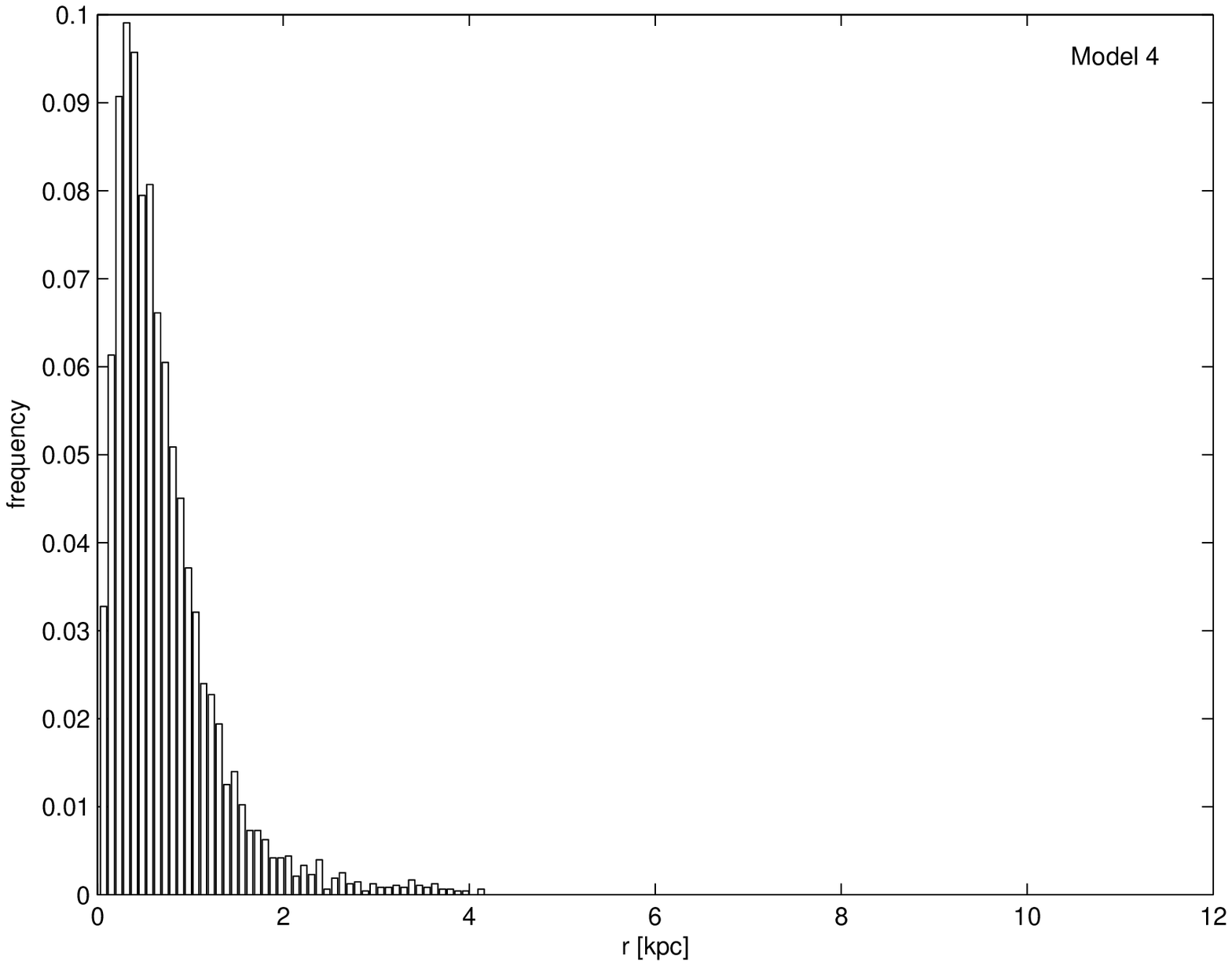}
%%\vspace{2cm}
\caption{Model 4: distribution of the gravitational wave frequency and distance from the Sun for detected sources with the advanced Virgo detector. \label{fgw_6_r_adv}}
\end{figure}
\begin{figure}
\includegraphics[width=84mm]{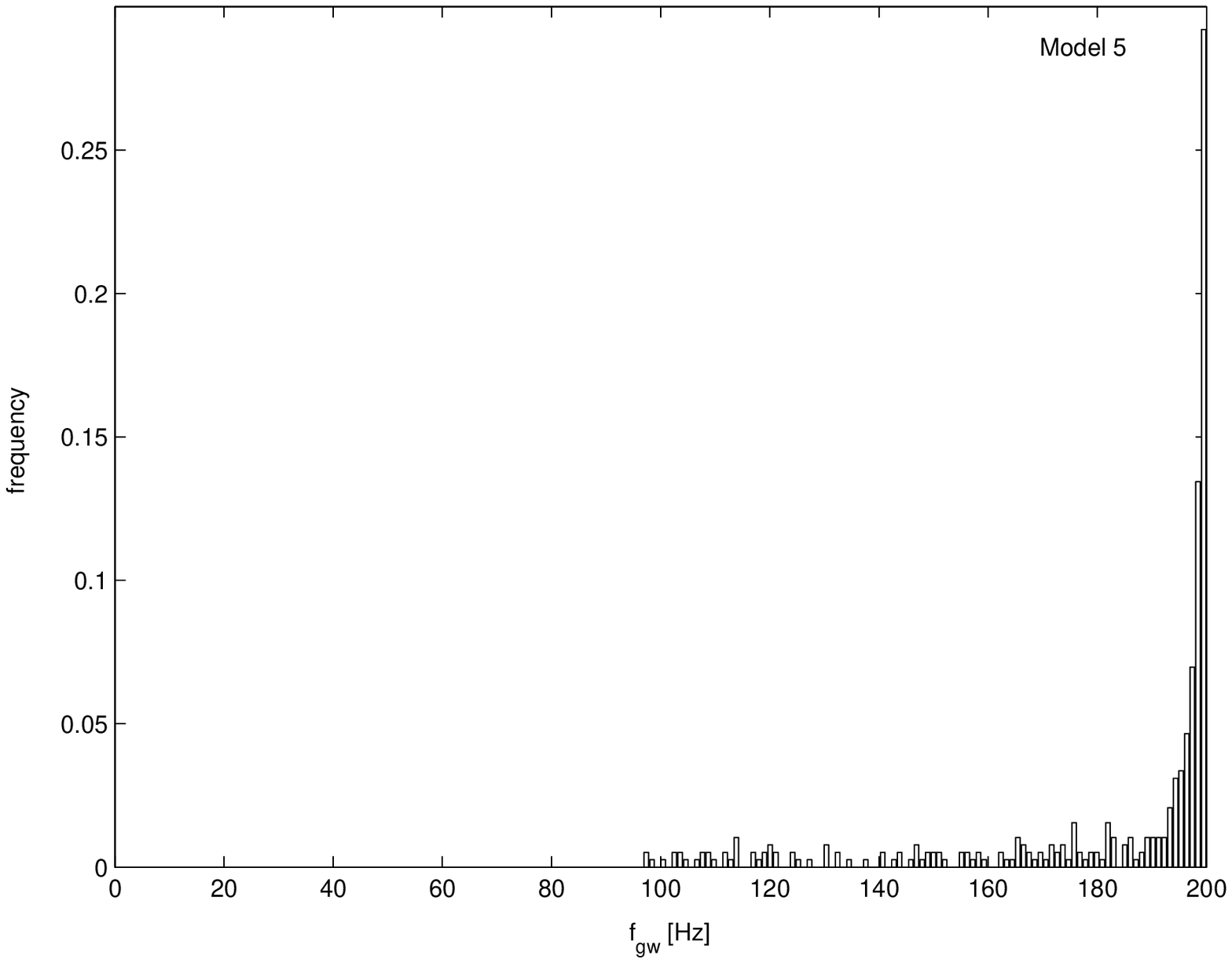}\includegraphics[width=84mm]{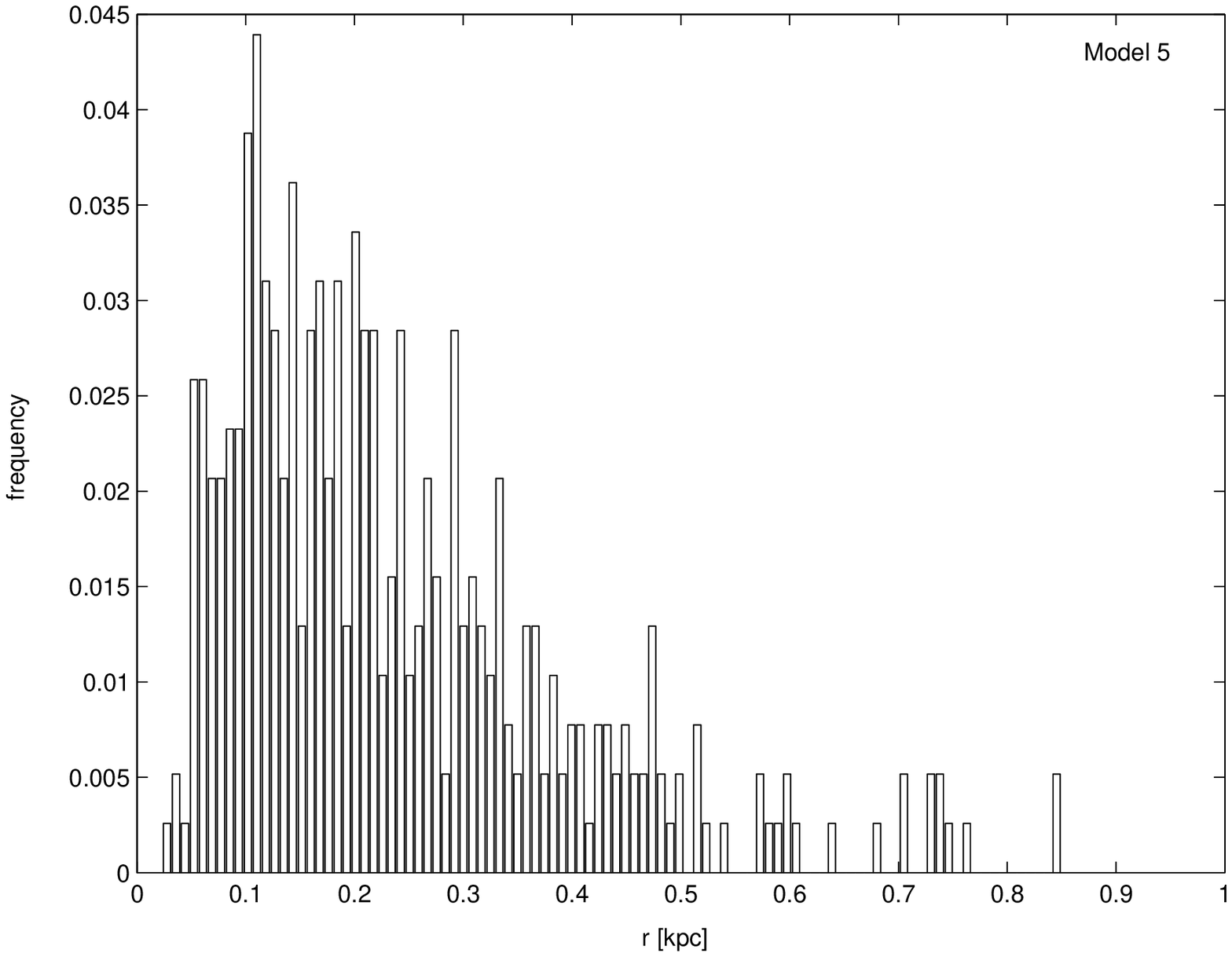}
%%\vspace{2cm}
\caption{Model 5: distribution of the gravitational wave frequency and distance from the Sun for detected sources with the advanced Virgo detector. \label{fgw_7_r_adv}}
\end{figure}
Fig.(\ref{fgw_6_r_adv}) and Fig.(\ref{fgw_7_r_adv}) refer respectively to Model 4 and Model 5. As already seen for the Virgo detector, for this kind of initial period distribution the 
detected sources have frequencies that tend to accumulate near the maximum possible value ($200~Hz$). This is particularly 
evident in Model 5 for which the detection rate would rapidly decrease if the maximum possible frequency were even a 
little less than $200~Hz$. No detection is expected for Model 6.
\begin{figure}
\includegraphics[width=84mm]{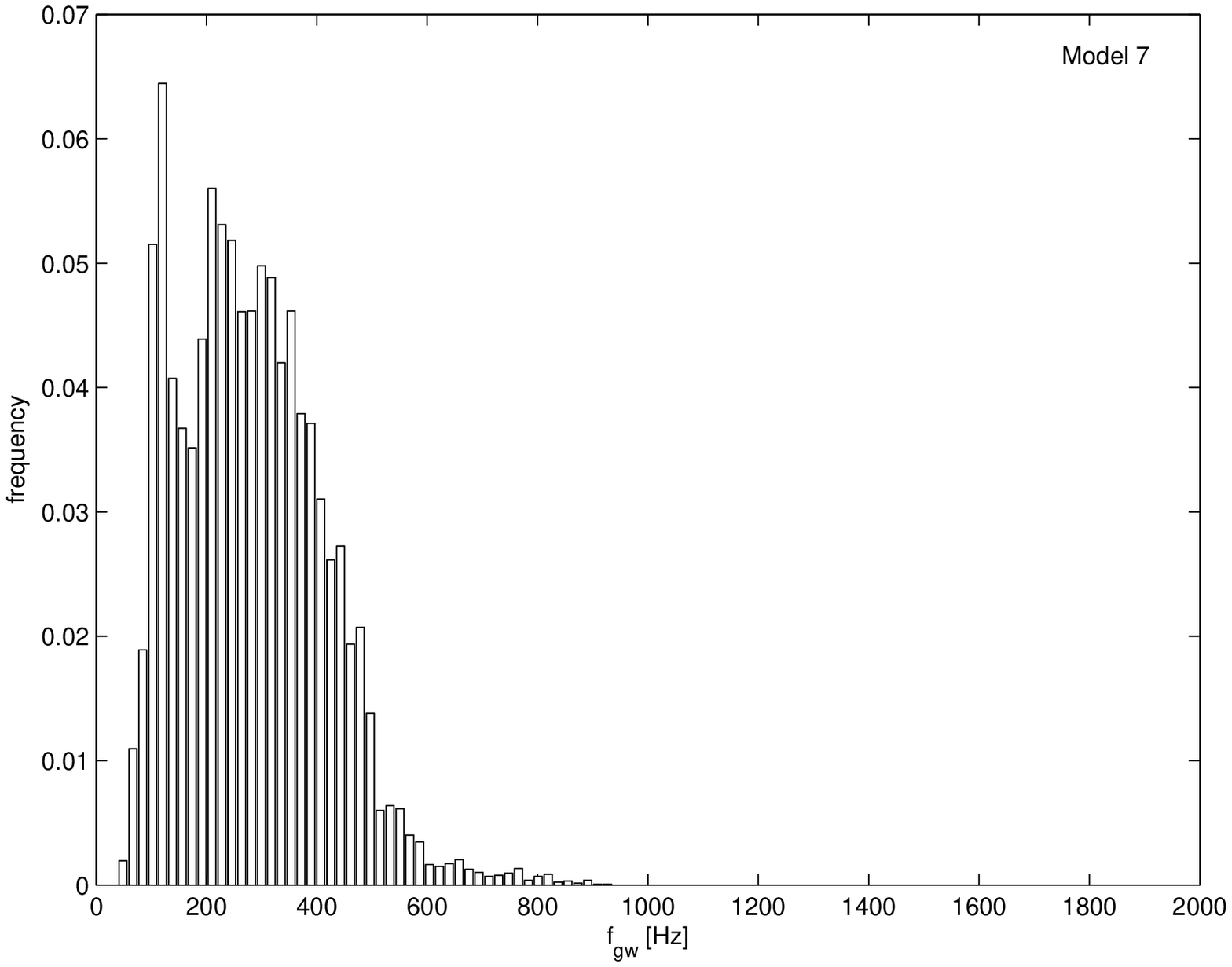}\includegraphics[width=84mm]{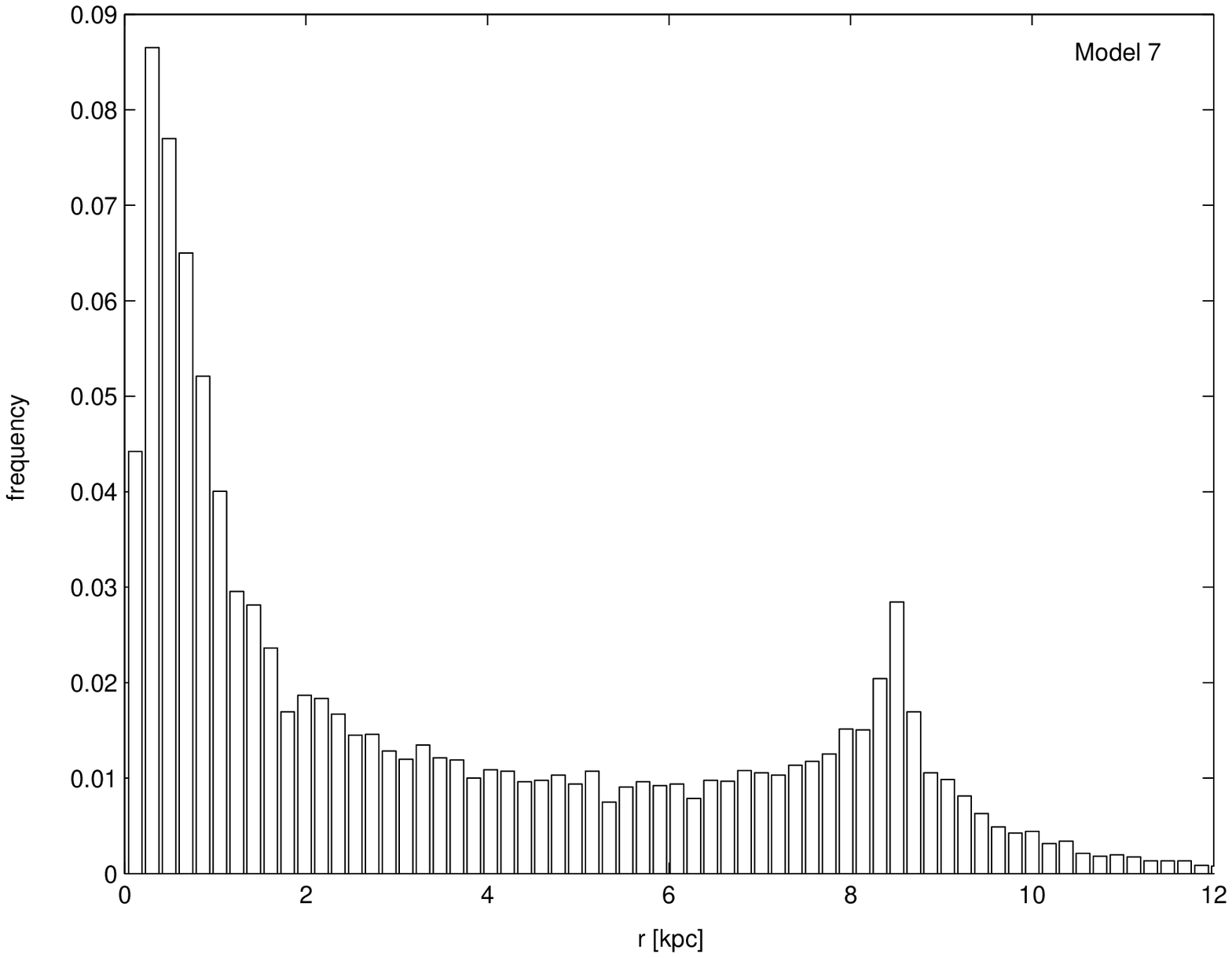}
%%\vspace{2cm}
\caption{Model 7: distribution of the gravitational wave frequency and distance from the Sun for detected sources with the advanced Virgo detector. \label{fgw_6_u_adv}}
\end{figure}
\begin{figure}
\includegraphics[width=84mm]{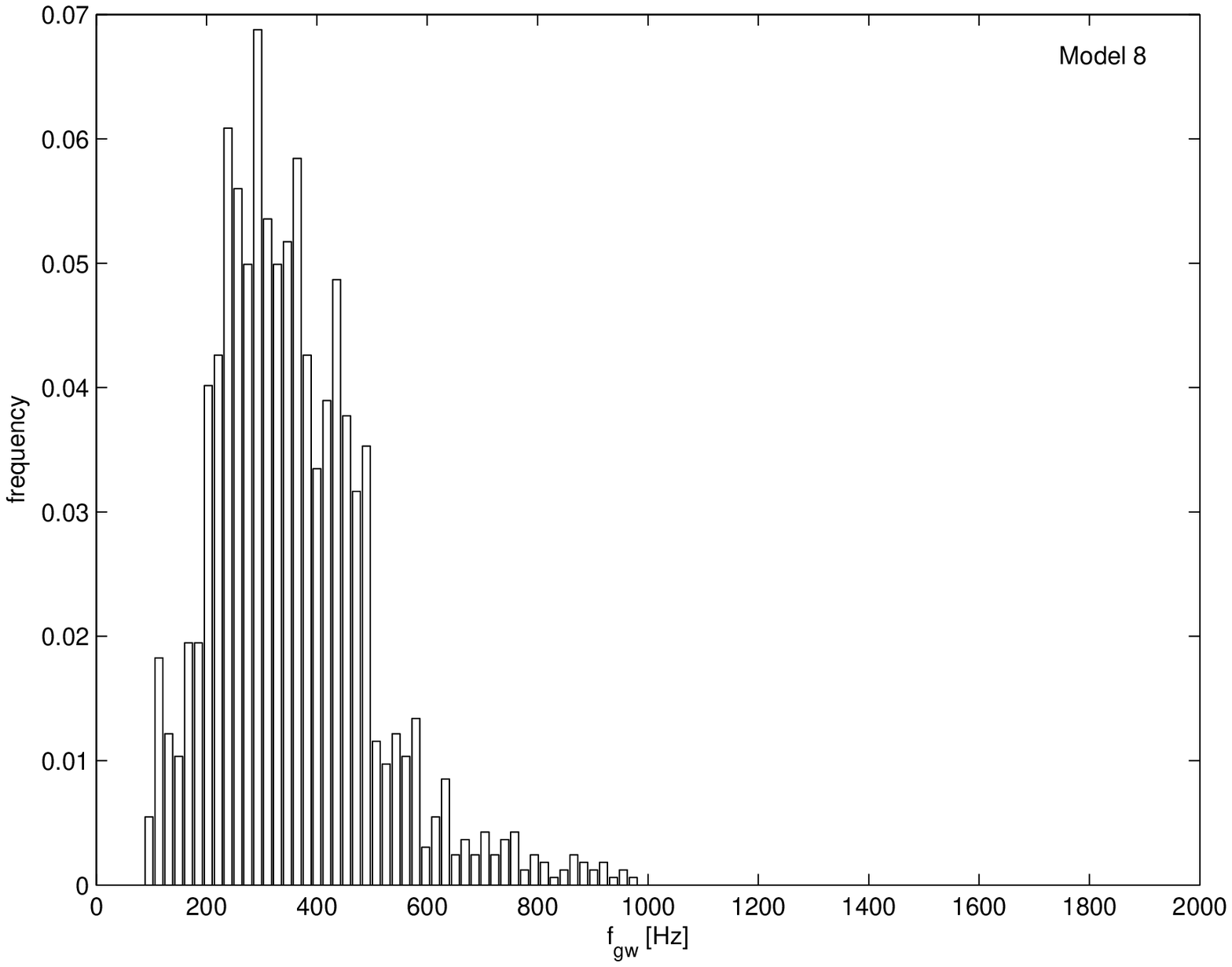}\includegraphics[width=84mm]{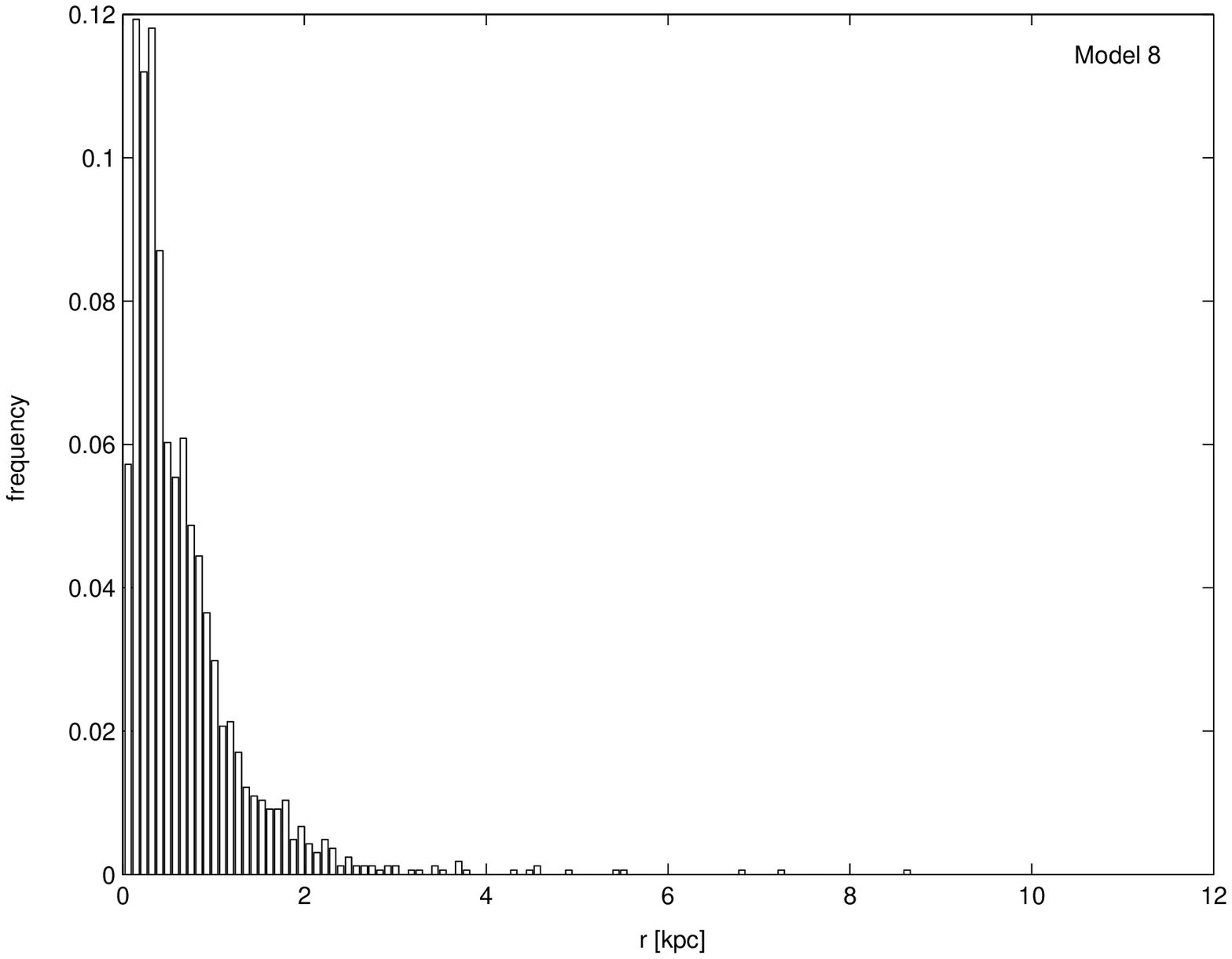}
%%\vspace{2cm}
\caption{Model 8: distribution of the gravitational wave frequency and distance from the Sun for detected sources with the advanced Virgo detector. \label{fgw_7_u_adv}}
\end{figure}
\begin{figure}
\includegraphics[width=84mm]{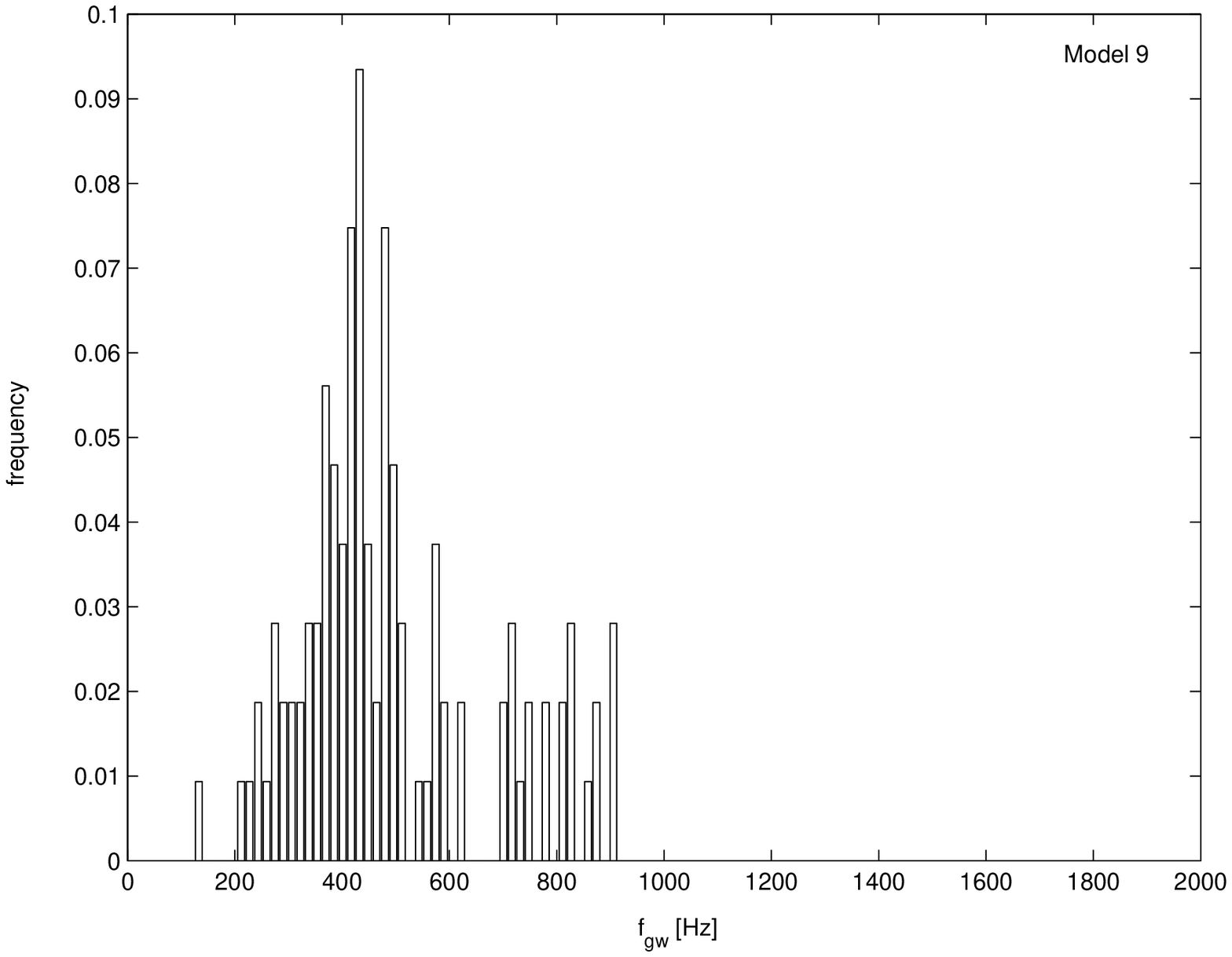}\includegraphics[width=84mm]{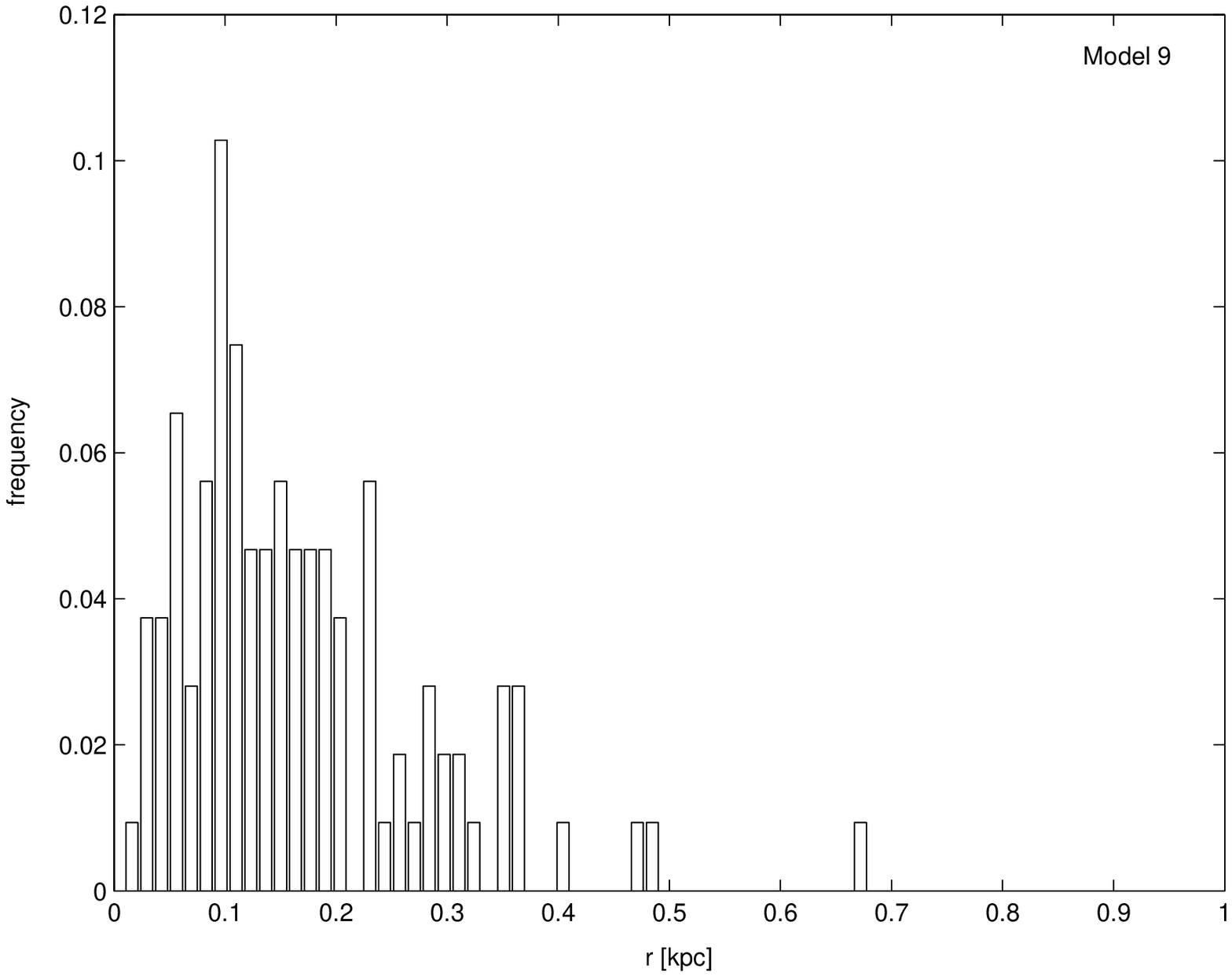}
%%\vspace{2cm}
\caption{Model 9: distribution of the gravitational wave frequency and distance from the Sun for detected sources with the advanced Virgo detector. \label{fgw_8_u_adv}}
\end{figure}
Figs.(\ref{fgw_6_u_adv},\ref{fgw_7_u_adv},\ref{fgw_8_u_adv}) refer, respectivley, to Models 7,8,9. For Model 7, as for Model 1, the detection distance arrives beyond the galactic centre and this produces the secondary peak in Fig.(\ref{fgw_6_u_adv}). Most of the detectable sources are near the galactic plane but, as for Model 1, the loudest sources are more spread: half of the $1\%$ strongest emitters has a declination, in modulus, larger 
than $16.6^o$. The distance distributions of Models 7-9 are rather similar to those we find for Models 1-3 while 
concerning the frequency distribution the main difference is that the maximum frequency is always less than $1000~Hz$, 
due to the minimum initial period equal to $2~ms$. 
\begin{table*}
%\begin{center}
\begin{minipage}{100mm}
\caption{For each kind of initial period distribution and for three different values of the fraction $\lambda$, in 
columns $2-4$ the minimum allowed value of the mean ellipticity $\overline{\epsilon}$ in order to have at least one detectable source is
reported; in column $5$ there is the minimum needed fraction if $\overline{\epsilon}=10^{-6}$. The advanced 
Virgo sensitivity shown in Figs.(\ref{pulsar1},\ref{pulsar2}) is assumed to hold.\label{tab5}}
\begin{tabular}{ccccc}
\hline
$p(P_0)$ & $\lambda=0.1$ & $\lambda=0.01$ & $\lambda=0.001$ & $\lambda_{min}$ \\
\hline
standard & $10^{-8}$ & $6\cdot 10^{-8}$ & $8.6\cdot 10^{-7}$ & $8.5\cdot 10^{-4}$ \\
r-modes &  $5.5\cdot 10^{-8}$ & $4.3\cdot 10^{-7}$ & $-$ & $4.2\cdot 10^{-3}$ \\
r-modes+fall-back & $1.9\cdot 10^{-8}$ & $1.2\cdot 10^{-7}$ & $-$ & $1.6\cdot 10^{-3}$ \\
\hline
\end{tabular}
%\end{center}
\end{minipage}
\end{table*}
In Tab.(\ref{tab5}) we have reported, for each initial period distribution function and for three different values of the fraction $\lambda$, 
the minimum value of the neutron star 
mean ellipticity needed in order to have at least one detectable source; the minimum allowed fraction $\lambda_{min}$, 
assuming $\overline{\epsilon}=10^{-6}$, is also shown. A value $\lambda=10\%$ would allow us to detect sources with mean 
ellipticity as small as $\sim 10^{-8}$. Even for $\lambda=10^{-3}$ we could still have a detection, if the mean ellipticity is $\overline{\epsilon}=8.6\cdot 10^{-7}$ and if the period distribution were described bt the "standard" model. If $\overline{\epsilon}=10^{-6}$, the minimum needed fraction varies between 
$8.5\cdot 10^{-4}$ and $4.2\cdot 10^{-3}$, depending on the initial period distribution.  

\section{Possible mechanisms allowing the existence of a population of GWDNS} \label{lowb}
GWDNS could be the result of an evolutionary path different from that of pulsars or could belong to 
the same population of pulsars, but with a very low dipolar magnetic field. Such a low magnetic field could be produced by 
several mechanisms some of which are now shortly discussed. 

According to \citet{gepp}, post core-collapse hypercritical accretion can submerge any initial magnetic field of the 
newborn neutron star, because the ram pressure at the star surface is much larger than the magnetic field pressure. Later 
diffusion of the magnetic field could produce a delayed switch-on of a pulsar, with timescale up to millions of years. 
There are, however, many possible factors which could reduce or even suppress the fall-back, like a high kick velocity or 
the injection of rotational energy in the  highly dirty environment of the new-born neutron star. It is not possible at 
present to estimate which fraction of newborn neutron stars undergoes a phase of hypercritical accretion. As most detected 
objects have, in our simulations, ages below $10~Myr$ (particularly in the case of the inital Virgo detector), this mechanism could be 
efficient in producing a population of GWDNS.

If the magnetic field of neutron stars is associated with the alignement of the magnetic moment of neutrons in the stellar 
crust, the orientation of the magnetic field will be influenced by the coupling between intrinsic spin and gravitational 
field of the star. This coupling induces the alignement of the magnetic axis respect to the rotation axis \citep{zhang}. 
In such a situation,  
\citet{zhang} have shown that the timescale of alignement is inversely porpoprtional to the magnetic field strength and 
proportional to the pulsar initial period. As a consequence, the inclination angles of weak field-short initial period 
pulsars should concentrate toward lower values.
The relation they find for the time evolution of the inclination angle $\alpha$ is
\begin{equation}
F[\alpha(t)]=F[\alpha(0)]-0.023\left({B_{in}\over {10^{12}~G}}\right)^{-1}\left({B\over {10^{12}~G}}\right)^{-2}\ln{\left({P(t)\over P_0}\right)}
\label{alphat}
\end{equation}
where $F[\alpha]=\alpha-{1\over 2}\sin{2\alpha}$ and $B_{in}$ is the star inner field.

A third possible mechanism is the decay of the neutron star magnetic field. 
As it is well known, the time evolution of the magnetic field in isolated neutron stars is still a controversial issue. 
A firm point is that the observation of radio pulsars rules out decay times less than $\sim 10~Myr$ but larger decay times
are 
still possible. The faster is the decay of the magnetic field and the larger is the fraction of 
GWDNS in the neutron star population. 

\section{Conclusions}
In this paper we have described the Monte Carlo simulation of a population of asymmetric neutron star which evolve through 
the emission of gravitational waves and have estimated the detectability of the emitted continuous gravitational signals. 
As far as we know, a quantitative study of this kind of sources has not been done before.
Some important features of this work are: neutron star ellipticity is not considered a constant, but distributed according to an 
exponential law; the presence of the Gould Belt is taken into account; the number of detectable sources is estimated 
assuming a realistic data analysis method is applied (in fact, the analysis method we will apply to the data of the Virgo 
detector, once they will be available). 
We have performed the simulation for values of the neutron star mean ellipticity between $10^{-6}$ and $10^{-8}$ and for 
three different models of the initial rotation period distribution corresponding, respectively, to a "standard" pulsar population, 
to a population of neutron stars spun-down by {\em r-modes} and to a population subject, after the supernova explosion, to the combined effect of {\em r-modes} and matter fall-back. The number of detections is estimated using the Virgo target sensitivity and a possible sensitivity curve of 
an advanced Virgo detector.
No large difference is expected for LIGO interferometers because very few sources are likely to be detected in the low frequency range, 
say below $40~Hz$, where the target sensitivity of Virgo is much better than that of LIGO. 
We do not have any clue of what the number of GWDNS could be, then the number of detectable sources is parametrized by the fraction $\lambda$ of the total neutron star population made of this kind of objects. 
For detectors of the first generation, we need a fraction $\lambda \sim 8\%$ to have one detectable source, for  $\overline{\epsilon}\sim 10^{-6}$. Smaller values, $\overline{\epsilon}\sim 2\div 5\cdot 10^{-7}$, require a fraction $\lambda$ as high as $\sim 50\%$.
The expected number of detections for advanced interferometers is about two orders of magnitude larger than that we find for initial detectors. This means that, assuming a value for $\lambda$, for advanced detectors the minimum detectable mean ellipticity is much smaller or, viceversa, for a given value of the mean ellipticity, the minimum needed fraction of neutron stars being GWDNS, in order to have at least one detectable source, is much lower. 
We find that a value of $\lambda$ as small as $0.001$ is enough to have one detectable source for high mean ellipticities, while if $\lambda \sim 10\%$, the minimum needed mean ellipticity is of the order of $10^{-8}$.   

In principle, GWDNS as old as the Galaxy could contribute to the detected population, because for many objects the slow 
gravitational spin-down does not bring the signal frequency below the detector low frequency cut-off even for age of 
the order of $10^{10}~yr$. However, in practice detectable sources are often very young, especially in the case of 
interferometers of the first generation, with ages of few Megayears or less. 
For many models, most detectable sources are very near to the Sun, a few hundreds parsec at most, and the Gould Belt 
contributes significantly to the detectable population. If these sources are really located in the solar neighbourhood, 
some of them could be detected in the electromagnetic band through their thermal emission. The distance "reached" by 
interferometers of the second generation is much larger than that of initial detectors: even sources located in the 
galactic centre can be detected, for some models, but strongest sources are expected to be near and then distributed rather homogeneously around the Sun. This, and the poor knowledge we have about typical frequencies and spin-down rates of GWDNS, are the reason why a {\em blind} search is needed in any case. As it is by now well known, a {\em blind} search can be performed only using non optimal data analysis methods with reduced sensitivity respect to optimal ones (typically a factor of 2-3) but which require a reasonable, although large, amount of computing resources. More precisely, the larger is the available computing power and the wider can be the search, covering a larger portion of the source parameter space.

This work has shown that GWDNS are potentially interesting sources of gravitational radiation. Clearly, the very poor knowledge we have about parameter values, in particular the ellipticity, prevents us from giving firm estimations of the expected number of detections. By the way, even for interferometers of the first generation, already taking data or being on the point to start data collection, it is possible, although rather unlikely, that some detection takes place. Possibilities of detection are obviously much better for interferometers of the second generation, which could be able to catch several sources. Indeed, gravitational wave astronomy would be the best tool to study the properties and demography of GWDNS.

\appendix

\section[]{Other mechanisms competing with the gravitational emission of GWDNS}
In the previous sections we have considered GWDNS as isolated asymmetric neutron stars with a magnetic field strength low 
enough so that the gravitational spin-down is dominant respect to the standard dipolar magnetic field spin-down. 
However, as already anticipated in Section 2, a neutron star can pass through different evolutionary stages in which 
the 
spin-down is due to other mechanisms, like those connected to the interaction of the star with the interstellar medium (ISM). In 
this section we shortly remind the standard classification of isolated neutron stars evolutionary stages and try to 
understand if mechanisms different from the dipolar field spin-down can overcome the gravitational spin-down in GWDNS.

According to the standard classification \citep{popov}, basically three different evolutionary stages can be outlined for 
neutron stars. In the {\em ejector} phase the gravitational energy density at the accretion radius is lower than the 
energy density of the momentum outflow produced by the rotating magnetic field; neglecting the possible emission of 
gravitational radiation, the spin-down is due to the rotating dipolar magnetic field. A neutron star in this phase can 
appear, but not necessarily, as a pulsar. This phase, depending on the strength of the field and on the star velocity, 
can last even for the whole neutron star life. 
The {\em propeller} phase starts when the gravitational energy density at the accretion radius is larger than the energy density due to the momentum outflow produced by the rotating magnetic field. This condition is verified when the rotation
period is larger than 
\begin{equation}
P_{crit}\sim 10\left({B\over {10^{12}G}}\right)^{1\over 2}\left({\dot{M}\over{10^8kgs^{-1}}}\right)^{-{1\over 4}}\left({r_{acc} \over {10^{12}m}}\right)^{1\over 8}\left({M\over M_{\sun}}\right)^{{1\over 8}}~s
\label{pcrit1}
\end{equation}
where 
\begin{equation}
r_{acc}\sim 3\cdot 10^{12}\left({M\over M_{\sun}}\right)\left({v\over {10kms^{-1}}}\right)^{-2}~~m
\label{racc}
\end{equation}
is the {\em accretion radius}.
The accretion rate on a neutron star, in the Bondi theory, is
\begin{equation}
\dot{M}\sim 10^{8}n\left({M\over M_{\sun}}\right)^2\left({v\over {10kms^{-1}}}\right)^{-3}~~kgs^{-1}
\label{dotM}
\end{equation}
where $n$ is the particle density in the ISM ($\sim 1~cm^{-3}$) and $v$ is the neutron star velocity.
Another characteristic length which we will use in the following is the {\em Alfv\`{e}n radius}
\begin{equation}
r_A\sim 2\cdot 10^8 \left({B\over {10^{12}G}}\right)^{4\over 7} \left({\dot{M}\over{10^8kgs^{-1}}}\right)^{-{2\over 7}}\left({R\over {10km}}\right)^{12\over 7}\left({M\over M_{\sun}}\right)^{-{1\over 7}}~m
\label{rA}
\end{equation}
In the {\em propeller} phase the neutron star can start to interact with the interstellar medium: matter can fall toward the neutron star up to the {\em Alfv\`{e}n radius} where the magnetic energy density balances the kinetic energy of the falling matter. The co-rotating magnetosphere will prevent the accretion of matter on the stellar surface, unless the gravitational acceleration at the {\em Alfv\`{e}n radius} is larger than the centrifugal acceleration {\it and} the {\em Alfv\`{e}n radius} is smaller then the accretion radius. The latter condition implies that the neutron star velocity satisifes the inequality
\begin{equation}
v<410n^{1\over {10}}\left({B\over {10^{12}G}}\right)^{-{1\over 5}}~~kms^{-1}
\label{vcrit}
\end{equation}
The former condition translates in a second critical period
\begin{equation}
P_A\sim 1000\left({B\over {10^{12}G}}\right)^{6\over 7} \left({\dot{M}\over{10^8kgs^{-1}}}\right)^{-{1\over 2}}\left({M\over M_{\sun}}\right)^{-{1\over 2}}~s
\label{pcrit2}
\end{equation}
For periods $P>P_A$ we are in the {\em accretor} phase where the matter can fall up to the stellar surface. Both in the {\em propeller} and in the {\em accretor} phase the neutron star spin-down is not given by Eq.(\ref{omegadotem}), which holds in the {\em ejector} phase, then the condition to have a GWDNS, given by Eq.(\ref{B}), must change.  
From Eqs.(\ref{pcrit1},\ref{dotM}) we have that the condition the magnetic field must satisfy in order to bring the neutron star in the {\em propeller} phase is
\begin{equation}
B<B_{crit,prop}\sim 7.6\cdot 10^5 n^{1\over 2}\left({M\over M_{\sun}}\right)^{1\over 2}\left({v\over {10kms^{-1}}}\right)^{-1}\left({f\over {100Hz}}\right)^{-2}~~G
\label{bcrit1}
\end{equation}
Here and in the following equations the neutron star is supposed to have a radius $R=10~km$.
Assuming the neutron star is in the {\em propeller} phase, let us see which is the further condition we need on the magnetic field so that the spin-down is still dominated by the emission of gravitational waves. In the {\em propeller} phase the neutron star looses its rotational energy to the ISM and spins-down at a rate \citep{popov3}
\begin{equation}
\dot{\Omega}_{prop}={{\xi \dot{M}r^2_{cor}}\over {I}}\left(\omega_k(r_A)-\Omega \right)
\label{omegadotprop}
\end{equation}
where
\begin{equation}
r_{cor}=\left({GM\over {\Omega^2}}\right)^{1\over 3}
\end{equation}
is the {\em corotation radius},
\begin{equation}
\omega_k(r_A)=\sqrt{{{GM}\over {r^3_A}}}
\end{equation}
is the local keplerian velocity at $r_A$ and $\xi$ is a numerical factor depending on the geometry of the accretion flow, which for spherical accretion, typical of isolated neutron stars, is $\sim 10^{-3}$.
Using Eq.(\ref{omegadotgw}) and Eq.(\ref{omegadotprop}) and imposing that  
$Y'={\dot{\Omega}_{gw}\over{\dot{\Omega}_{prop}}}>1$ we find:
\begin{equation}
B>B'_{crit,prop}=8.9\cdot 10^{9}{{n^{1\over 2}\left({M\over M_{\sun}}\right)^{11\over 6}\left({v\over {10kms^{-1}}}\right)^{-{3\over 2}}\left({f\over {100Hz}}\right)^{-{7\over 6}}}\over{\left(1+7.1\cdot 10^5\xi ^{-1}\left({M\over M_{\sun}}\right)^{-{8\over 3}}\left({\epsilon \over{10^{-6}}}\right)^2
\left({v\over {10kms^{-1}}}\right)^{3}\left({f\over {100Hz}}\right)^{16\over 3}\right)^{7\over 6}}}~~G
\label{b2crit}
\end{equation}
For instance, assuming $\epsilon=10^{-6},~v=10~kms^{-1}, f=25~Hz$ we have $B_{crit,prop}\simeq 10^7~G$ and $B'_{crit,prop}\simeq 6\cdot 10^4~G$, that is, even if the magnetic field is so low that the neutron star moves to the {\em propeller} phase, the gravitational spin-down is still dominating, unless the magnetic field is nearly vanishing. This conclusion is even strenghtened if we consider that most neutron stars have typical velocity much larger than $10~kms^{-1}$ and that the accretion rate given by Eq.(\ref{dotM}) is only an upper limit and is probably much lower in real situations. 
From Eqs.(\ref{pcrit2},\ref{dotM}) we obtain also the condition on the magnetic field in order to enter in the {\em accretor} phase:
\begin{equation}
B<B_{crit,acc}\sim 1.5\cdot 10^6\left({M\over M_{\sun}}\right)^{7\over 4}\left({v\over {10kms^{-1}}}\right)^{-{7\over 4}}\left({f\over {100Hz}}\right)^{-{7\over 6}}~~G
\label{bcritacc}
\end{equation}
If this condition is satisfied, Eq.(\ref{vcrit}), which must be also satisfied to be in the {\em accretor} regime, gives
$v<6\cdot 10^3~kms^{-1}$, and then is also verified. 
Now we compare the spin-down rate of a neutron star in the {\em accretor} stage with that due to gravitational radiation.
The spin-down rate in the accretor stage can be written as \citep{kone}
\begin{equation}
\dot{\Omega}_{acc}=-k_t{{B^2R^6}\over r^3_{cor}}+K_{turb}
\label{omegadotacc}
\end{equation}
where the first term in the righ-hand side describes the braking moment of the forces due to the possible turbulization of the ISM, $k_t$ is a dimensionless costant with value $\sim 1$ and $K_{turb}$ is a random term which superimposes to the first one and can produce, on small time scales, both spin-down and spin-up of the neutron star.
Imposing that $Y''={\dot{\Omega}_{gw}\over{\dot{\Omega}_{acc}}}>1$ we find our last condition on the magnetic field strength in order to have a neutron star dominated by the emission of gravitational waves:
\begin{equation}
B<B'_{crit,acc}\sim 2.7\cdot 10^8 \left({M\over M_{\sun}}\right)^{1\over 2}\left({\epsilon \over{10^{-6}}}\right)\left({f\over {100Hz}}\right)^{3\over 2}~~G
\label{bcrit2acc}
\end{equation}
If Eq.(\ref{bcritacc}) is verified also Eq.(\ref{bcrit2acc}) is satisfied, at least if we limit ourselves to consider the frequency range where detectable (in principle) sources can be placed, that is $f>25~Hz$.
As a conclusion, we can say that, unless the magnetic field is nearly vanishing, a neutron star which is not in the {\em ejector} phase will be dominated by the emission of gravitational radiation.

\section*{Acknowledgments}

I want to thank Sergio Frasca and Fulvio Ricci for the stimulating discussions. I am also grateful to Sergio Frasca for 
having suggested the use of the principle of maximum entropy and to Michele Punturo for having provided a possible 
sensitivity curve of an advanced Virgo detector. I thank people working in the Tier-1 computing 
centre at CNAF (Bologna) and in the INFN-GRID project for the support they gave me in running the simulations on the 
INFN Production Grid. Finally, I am grateful to the anonymous referee for the constructive comments and suggestions aiming
at improving the quality of the paper. 

\label{lastpage}
\end{document}